\documentclass{article}
\usepackage{pifont}
\usepackage{nomencl}

\usepackage{natbib}
\bibpunct[, ]{(}{)}{,}{a}{}{,}%
\def\bibsep{\smallskipamount}%
\def\newblock{\ }%

\usepackage{graphicx}
\usepackage{amsmath}
\usepackage{amstext}
\usepackage[font=bf,skip=0.333\baselineskip]{caption}
\usepackage{xcolor}

\usepackage{wasysym} 
\usepackage{pdflscape}
\usepackage{physics}
\usepackage{mathdots}
\usepackage{yhmath}
\usepackage{cancel}
\usepackage[group-separator=\text{,},group-minimum-digits={3}]{siunitx}
\usepackage{afterpage}
\usepackage{array}
\usepackage{float}
\usepackage{enumitem} 
\usepackage{mathtools}
\usepackage{longtable}
\usepackage{multirow}
\usepackage{amssymb} 
\usepackage{multicol}
\usepackage{gensymb}
\usepackage{soul}
\usepackage{relsize}
\usepackage{tabularx}
\usepackage{hyphenat}
\usepackage[thinlines]{easytable}
\usepackage{booktabs}
\usepackage{adjustbox}
\usepackage{comment}
\usepackage{textcomp}
\usepackage{subcaption}
\usepackage{appendix}
\usepackage{rotating}
\usepackage{siunitx}

\newcounter{tableeqn}[table]
\renewcommand{\thetableeqn}{\thetable.\arabic{tableeqn}}
\newcounter{tablesubeqn}[tableeqn]

\usepackage[colorlinks,hidelinks, breaklinks]{hyperref}
\usepackage{ifthen}
\usepackage{url}

\usepackage{tikz}

\DeclareMathSizes{5}{5}{5}{5}

\makenomenclature
\renewcommand{\nomgroup}[1]{%
\ifthenelse{\equal{#1}{A}}{\item[\textbf{Sets}]}{%
\ifthenelse{\equal{#1}{B}}{\item[\textbf{Input parameters}]}{%
\ifthenelse{\equal{#1}{C}}{\item[\textbf{Decision Variables}]}{}}}
}
\usepackage[normalem]{ulem}
\DeclareRobustCommand{\mysout}[1]{%
  \LetLtxMacro\origcite\cite%
  \LetLtxMacro\origcitet\citet%
  \LetLtxMacro\origcitep\citep%
  \renewcommand{\cite}[2][]{\mbox{\origcite[##1]{##2}}}%
  \renewcommand{\citet}[2][]{\mbox{\origcitet[##1]{##2}}}%
  \renewcommand{\citep}[2][]{\mbox{\origcitep[##1]{##2}}}%
  \texorpdfstring{\sout{#1}}{#1}}
\newcommand{\SuggestEdit}[3][red]{{\color{#1}\mysout{#2}}{\color{#1}#3}}
\let\svthefootnote\thefootnote
\newcommand\colorfootnote[2][black]{\def\thefootnote{\color{#1}\svthefootnote}%
  \footnote{\color{#1}#2}\def\thefootnote{\color{black}\svthefootnote}}
\newcommand\RevComment[3][red]{\protect\colorfootnote[#1]{{\textbf{[#2: #3]}}}}

\newcommand\newrevisor[2]{
  \colorlet{#1}{#2}
  \expandafter\DeclareRobustCommand\csname#1\endcsname[2]{\SuggestEdit[#1]{##1}{##2}}
  \uppercase{\expandafter\DeclareRobustCommand\csname#1}\endcsname[1]{\RevComment[#1]{\MakeUppercase{#1}}{##1}}
}
\newcommand\hiderevisor[1]{
  \expandafter\renewcommand\csname#1\endcsname[2]{}
  \uppercase{\expandafter\renewcommand\csname#1}\endcsname[1]{}%
}

\definecolor{red}{RGB}{255,187,177}
\definecolor{ForestGreen}{RGB}{155,222,224}
\definecolor{blue}{RGB}{181, 164, 139}
\definecolor{yellow}{RGB}{199, 139, 201}
\definecolor{purple}{RGB}{13, 78, 79}
\definecolor{gray}{RGB}{128, 128, 128}

\definecolor{darkred}{RGB}{178, 34, 34}
\definecolor{darkbrown}{RGB}{101, 67, 33}
\definecolor{darkpink}{RGB}{151, 7, 68}
\definecolor{darkpurple}{RGB}{67, 33, 101}

\usepackage{tikz}
\usepackage{array}

\newcommand{\drawBoxHSC}[3]
{
  \begin{tikzpicture}
  \def\w{1} 
  \def\x{#1/100*\w} 
  \def\xl{#1/10000*\w} 
  \def\xu{#1/100*\w} 
  \filldraw[fill=red] (0,0) rectangle (\x,0.2); 
  \draw [gray] (0,0) rectangle (\w,0.2); 
  \draw (\x,0.1) -- (\xu,0.1) -- (\xu,0.15) -- (\xu,0.05);
  \end{tikzpicture} 
}

\newcommand{\drawBoxPM}[3]
{
  \begin{tikzpicture}
  \def\w{1} 
  \def\x{#1/100*\w} 
  \def\xu{#1/100*\w} 
  \filldraw[fill=purple] (0,0) rectangle (\x,0.2); 
  \draw [gray] (0,0) rectangle (\w,0.2); 
  \draw (\x,0.1) -- (\xu,0.1) -- (\xu,0.15) -- (\xu,0.05);
  \end{tikzpicture} 
}

\newcommand{\drawBoxSHOB}[3]
{
  \begin{tikzpicture}
  \def\w{1} 
  \def\x{#1/100*\w} 
  \def\xu{#1/100*\w} 
  \filldraw[fill=ForestGreen] (0,0) rectangle (\x,0.2); 
  \draw [gray] (0,0) rectangle (\w,0.2); 
  \draw [gray] (0,0) rectangle (\w,0.2); 
  \draw (\x,0.1) -- (\xu,0.1) -- (\xu,0.15) -- (\xu,0.05);
  \end{tikzpicture} 
}

\newcommand{\drawBoxSHOO}[3]
{
  \begin{tikzpicture}
  \def\w{1} 
  \def\x{#1/100*\w} 
   \def\xu{#1/100*\w} 
  \filldraw[fill=blue] (0,0) rectangle (\x,0.2);
  \draw [gray] (0,0) rectangle (\w,0.2); 
  \draw [gray] (0,0) rectangle (\w,0.2); 
  \draw (\x,0.1) -- (\xu,0.1) -- (\xu,0.15) -- (\xu,0.05);
  \end{tikzpicture} 
}

\newcommand{\drawBoxVSC}[3]
{
  \begin{tikzpicture}
  \def\w{1} 
  \def\x{#1/100*\w} 
  \def\xu{#1/100*\w} 
  \filldraw[fill=yellow] (0,0) rectangle (\x,0.2); 
  \draw [gray] (0,0) rectangle (\w,0.2); 
  \draw [gray] (0,0) rectangle (\w,0.2); 
  \draw (\x,0.1) -- (\xu,0.1) -- (\xu,0.15) -- (\xu,0.05);
  \end{tikzpicture} 
}

\usepackage{arxiv}

\usepackage[utf8]{inputenc} 
\usepackage[T1]{fontenc}    
\usepackage{hyperref}       
\usepackage{url}            
\usepackage{booktabs}       
\usepackage{amsfonts}       
\usepackage{nicefrac}       
\usepackage{microtype}      
\usepackage{lipsum}		
\usepackage{graphicx}
\usepackage{natbib}
\usepackage{doi}

\title{Managing Manufacturing and Delivery of Personalised Medicine: Current and Future Models}

\author{ \href{https://orcid.org/0000-0001-6837-6272}{\includegraphics[scale=0.06]{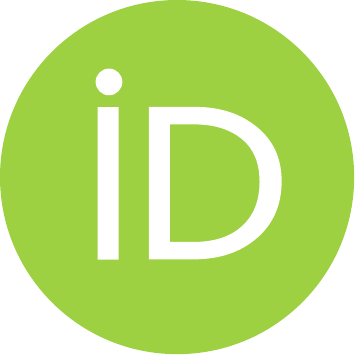}\hspace{1mm}Andreea Avramescu} \\
	Alliance Manchester Business School\\
	University of Manchester\\
	Manchester, UK \\
	\texttt{andreea.avramescu@manchester.ac.uk} \\
	\And
	\href{https://orcid.org/0000-0003-1236-3143}{\includegraphics[scale=0.06]{orcid.pdf}\hspace{1mm}Richard Allmendinger} \\
	Alliance Manchester Business School\\
	University of Manchester\\
	Manchester, UK \\
	\texttt{richard.allmendinger@manchester.ac.uk} \\
	\And
	\href{https://orcid.org/0000-0001-9974-1295}{\includegraphics[scale=0.06]{orcid.pdf}\hspace{1mm}Manuel López-Ibáñez} \\
	Alliance Manchester Business School\\
	University of Manchester\\
	Manchester, UK \\
	\texttt{manuel.lopez-ibanez@manchester.ac.uk} \\
}

\date{}
\renewcommand{\headeright}{}
\renewcommand{\undertitle}{}
\renewcommand{\shorttitle}{Managing Manufacturing and Delivery of Personalised Medicine}

\hypersetup{
pdftitle={Managing Manufacturing and Delivery of Personalised Medicine: Current and Future Models},
pdfsubject={math.OC},
pdfauthor={Andreea Avramescu, Richard Allmendinger, Manuel López-Ibáñez},
pdfkeywords={healthcare, supply chain, personalised medicine, mathematical model},
}

\begin{document}
\maketitle
\vspace{-2em}
\begin{abstract}
	  With almost 50\% of annual commercial drug approvals being Personalised Medicine (PM) and its huge potential to improve quality of life, this emerging medical sector has received increased attention from the industry and medical research, driven by health and care services, and us, the patients. Notwithstanding the power of Advanced Therapy Medicinal Products (ATMPs) to treat progressive illnesses and rare genetic conditions, their delivery on large scale is still problematic. The biopharmaceutical companies are currently struggling to meet timely delivery and, given high prices of up to \$2 million per patient, prove the cost-effectiveness of their ATMP.  The fragility of ATMPs combined with the impossibility for replacements due to the nature of the treatment and the advanced stages of the patient’s condition are some of the bottlenecks added to a generally critical supply chain. As a consequence, ATMPs are currently used in most cases only as a last resort. ATMPs are at the intersection of multiple healthcare logistic networks and, due to their novelty, research around their  commercialisation is still in its infancy from an operations research perspective. To accelerate technology adoption in this domain, we characterize pertinent practical challenges in a PM supply chain and then capture them in a holistic mathematical model ready for optimisation. The identified challenges and derived model will be contrasted with literature of related supply chains in terms of model formulations and suitable optimisation methods. Finally, needed technological advancements are discussed to pave the way to affordable commercialisation of PM.
  
\end{abstract}


\section{Introduction} \label{sec:introduction}

It is widely accepted that traditional pharmaceuticals are efficient for approximately 60\% of the population only \citep{ermak_2016}. The ``one-formula-fits-all'' drugs and the currently reactive healthcare approach are following a trial-and-error prescription of medicines by having its primary focus on the disease rather than the patient \citep{agyeman_ofori_asenso_2015}. However, the past decades have seen increased interest in understanding patient variability. New evidence is being accumulated supporting the statement that the organism's response to drugs is driven by genetic, environmental, and social conditions \citep{vogenberg_et_al_2010}. Among other factors, it was this evolution of the pharmacokinetics (``what the body does to the drug'') and pharmacodynamics (``what the drug does to the body'') that allowed medical researchers to discuss new treatments and develop dedicated approaches known under the umbrella term of Personalised Medicine (PM) \citep{morse_kim_2015}. With a thriving popularity in an industry that exceeds \$250 billion worldwide \citep{moorkens_2017}, the advancement of innovative therapies in the medical sector has been rapid. It is estimated that by 2030, approximately \num{50000} people could be treated yearly with over 60 approved treatments \citep{quinn_et_al_2019}.

Notwithstanding the potential of PM to treat and even cure patients where all other forms of treatment have failed, the number of individuals that had benefited from a personalised therapy is negligible \citep{meij_et_al_2019}. The delivery and development of PM treatments is still problematic. The current state of play is that simulation and optimisation of PM delivery relies heavily on  models borrowed from the biopharmaceutical field. The PM therapies, known as personalised Advanced Therapy Medicinal Products (ATMPs), are however complex and driven by a different set of constraints, forcing the pharmaceutical industry to rethink the development, manufacturing, and delivery of the products from a continuous, off-the-shelf approach to a batch and on-demand model \citep{trainor_et_al_2014}. Without significant manufacturing and supply innovations in both the physical and digital space, the promise of targeted healthcare will remain accessible only on a small scale \citep{elverum_whitman_2019}. The novelty of the field, from an operations research perspective, calls for extensive research and the implementation of holistic models.

ATMPs could be formally placed at the junction of multiple supply chains: those that support biopharmaceutical products and those that support substances of human origin (SHO), namely blood transfusions and organ transplantation \citep{rutherford_et_al_2017}. Critical concerns in the ATMPs, such as shelf life, patients stratification, and uncertain supply and demand are also some of the principal bottlenecks of the SHOs supply chains. However, ATMPs have a distinctly fragile composition and in most cases are impossible to replace due to the advanced stage of a patient’s disease. In addition, the responsive nature of PM acknowledges each person's characteristics as a clear indicator of the way the treatment should be applied \citep{chouchane_et_al_2011}. In this sense, the patient has increasingly started to have an active role in the treatment and is now an integral factor of the supply chain. Unlike the more common biopharmaceuticals, in a personalised ATMP scenario individual's health condition is directly influencing the scheduling of the product, with the patient becoming one of the suppliers of raw materials and the main customer of the final product. All these lead to high costs and complex manufacturing processes, forcing these medicines to be used only as a last resort.

Despite its differences with other mature healthcare supply chains, ATMPs can also benefit from operations research and management perspectives by developing models capable of aiding in the optimisation of the products. Hence, this paper presents the most common challenges faced by the ATMP supply chain. The need for extensive reviews that incorporates literature from different fields to understand the potential impact of healthcare operations management has also been highlighted as a research priority by \citet{singh_et_al_2020}. Additionally, we propose a novel mathematical formulation for personalised ATMPs based on extensive literature concerned with mathematical modelling and solution methods previously used in the related healthcare supply chains. The formulation will target strategical and operational supply chain levels of personalised ATMPs, integrating multi-type facility location-allocation problems with a multi-mode manufacturing process.

The rest of the paper is structured as follows. Sections~\ref{sec:PM_landscape} and \ref{sec:integration_healthcare} describe the current PM landscape and its supply chain. The methodology for the systematic review is outlined in Section~\ref{sec:methodology}. The PM supply chain and its general integration in the wider field of healthcare supply chains is described in Section~\ref{subsec:integration_healthcare}, while the complete literature review is presented in Section~\ref{sec:literature_review}. Drawing from this analysis, a mathematical model is presented in Section~\ref{sec:math_models}. The paper ends with conclusion and directions for future research in Section~\ref{sec:conclusion}. 

    \section{Personalised Medicine Landscape} \label{sec:PM_landscape}

By 2020 there were 42 ATMPs approved to market \citep{eder_wild_2019} and over \num{2000} ongoing clinical trials in Europe alone \citep{AfRM_2019}, becoming one of the fastest growing areas of the pharmaceutical industry. The rapid progression is unlikely to cease in the foreseeable future given the ability of the novel therapies to cure rare and genetic conditions. Namely, there are only 8\% of these orphan diseases---i.e. a disease that affects approximately less than \num{200000} people worldwide \citep{aronson_2006}---that had at least one drug approved at the beginning of 2018 \citep{seoane_vazquez_et_al_2019}.

The shift towards personalisation has important implications not only for patients and healthcare executives \citep{betcheva_et_al_2020}, but also for the general population. From a social sustainability perspective, personalised therapies have the potential to minimise the growing resistance of certain diseases to antibiotics and can potentially lead to their eradication \citep{moser_et_al_2019}. From an environmental perspective, the ATMPs can lead the generally unsustainable pharmaceutical supply chains towards zero waste. As a result, breakthrough ATMPs have not only attracted the interest of researchers and scientists, but also the public’s and government's attention. Through investments in projects such as The Precision Medicine Initiative in the USA \citep{collins_varmus_2015} and HORIZON2020 in Europe \citep{nimmesgern_et_al_2017}, PM has become one of the healthcare research priorities. 

However, the current approach for the commercialisation of ATMPs has led to major challenges for some companies, such as the cases of Provenge \citep{jaroslawski_toumi_2015} and ChondroCelect \citep{abou_et_al_2016}, or to lower quality treatments, such as the case of Kymriah \citep{bersenev_kili_2018}. The ATMP's are among the most expensive medical treatments. Their price tag is a reflection of (i)~the high development cost through the clinical stages and up to commercialisation and (ii)~the high manufacturing and delivery costs, given a low global demand and the need for cold chain logistics, i.e., temperature-controlled supply chain \citep{abou_et_al_2016}.

\section{ATMPs Supply Chain Configuration} \label{sec:integration_healthcare}

\begin{figure}[H]
    \centering
    \includegraphics[width=0.9\textwidth]{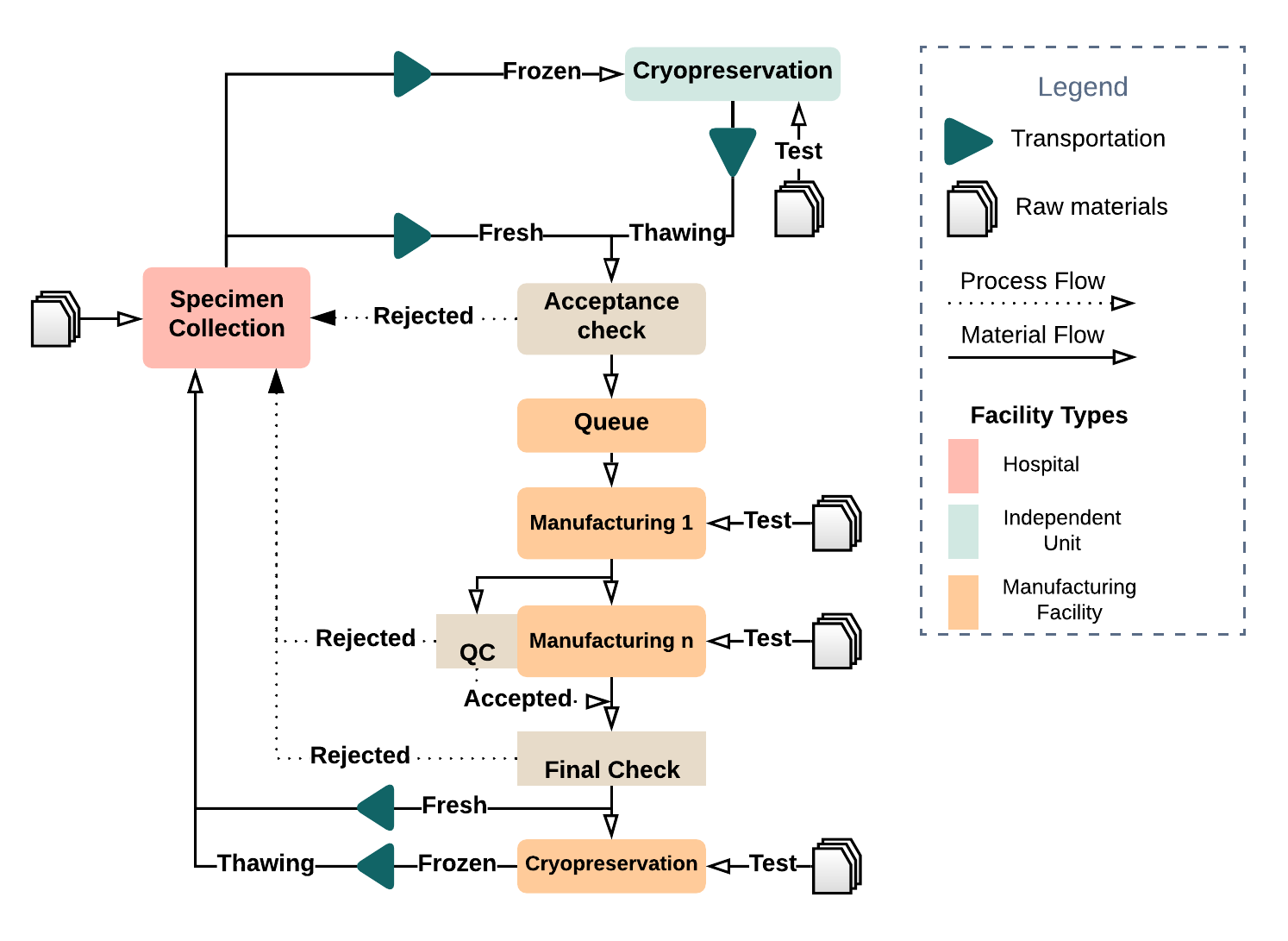}
    \caption{Schematic flowchart of the autologous supply chain for an ATMP, including independent cryopreservation unit, manufacturing steps, and failure possibility.}
    \label{fig:supply_chain_general}
\end{figure}

The generic supply chain configuration for autologous ATMPs is graphically presented in Figure~\ref{fig:supply_chain_general}. It starts with material collected from the donor (allogeneic process) or the patient (autologous process) through an apheresis procedure, for products that have blood as starting material, or a biopsy for products using solid tumours. Through leukapheresis, components, such as plasma or white blood cells, are separated and stored in a special container, while the remainder of the blood is returned to the patient. These procedures can only be conducted at candidate hospitals that have obtained FACT (Foundation for the Accreditation of Cellular Therapy) approval. The collected product is then transported to a manufacturing facility where it is genetically modified \citep{portner_et_al_2017}. Hence, finding optimal locations for the manufacturing facilities to optimise the delivery of the products is a first step in the supply chain network design. 

Facility location problems have been studied extensively in the literature, however the ATMP delivery is additionally challenged by the products sensitivity to temperature variations. Using living cells, the entire process from this moment and until the ATMP is returned to the patient, has a short shelf-life; from a few hours to a few days depending on the type of product. Therefore, transportation is commonly done within a cold supply chain between $-60$\degree C and cryogenic temperatures. Freezing the material extends the preservation timeframe and allows for off-the-shelf distribution, especially important for allogeneic products, relaxing altogether the time constraints \citep{rafiq_et_al_2017}. The cryopreservation cannot be conducted at the hospital and is currently undertaken at an independent facility. In an ATMP network, it is therefore important to optimise the locations of both manufacturing and cryopreservation facilities at the same time. 

A cryopreserved product needs to be thawed before starting the manufacturing process and before the final administration at the hospital. This process alongside poor logistics increases the risk of damaging the ATMP \citep{woods_et_al_2016}. Maintaining the product viable is crucial as the possibility of replacements is minimal. The autologous process shown in Fig.~\ref{fig:supply_chain_general} involves patients with advanced-stage diseases and their poor health condition might preclude another apheresis procedure. In addition to thawing associated risks, the failure rate of a product is also influenced by the way each step of the manufacturing process is organized. The processing tasks and their duration depend on the product but are mainly classified into four categories: cell and gene therapy, cell therapy, gene therapy, and tissue engineering. The manufacturing is currently labour-intensive and time-consuming, leading to a high cost of goods (COGs) and a long processing time per product \citep{lipsitz_et_al_2017}. However, various automation modes of production are currently available which can lower each task duration, ensuring a better standardisation between ATMPs, lower failure rates \citet{lopes_et_al_2020}, and relaxing the dependence on labour availability. Finding skilled workers in some geographical areas is challenging for the ATMP supply chain given the emerging highly specialised systems they are working with. The optimisation of the manufacturing process alongside facility location could then lead to a more resilient and efficient supply chain design. 

Each step of the manufacturing process is also influenced by various exogenous factors that can lead to lower quality therapies and ultimately become inefficient \citep{lipsitz_et_al_2016}. One such factor that is specific to a personalised supply chain is the patient’s health condition. As the starting material directly involves the patients, their health condition can influence the quality of the ATMP, and the number of blood samples required to obtain a high-quality final product. The patient becomes for the first time in the literature an integral factor of the network and drives the entire scheduling of an ATMP. Finally, once the manufacturing process is completed, the ATMPs are transported to the patient, either fresh or frozen. At this step, cryopreservation takes place within the manufacturing facility and not at an independent facility, but still carries the risks mentioned above. The product is administered only if the patient’s health condition allows it. 

    \section{Methodology} \label{sec:methodology}
The supply chain management (SCM) field has been defined in different ways  \citep{swanson_et_al_2018}, but it mainly refers to the production flow from start to end of a product with the aim of maximising profit, customer experience, and quality. This review will only consider papers related to the subfields of SCM that are critical for the PM supply chain \citep{rutherford_et_al_2017}, on strategical, tactical, and operational levels. Specifically, these include facility location, logistics and risk management; manufacturing and service processes; demand management and inventory planning; and service and patient scheduling. 

To minimise the double review of articles that have been considered in past studies, we have restricted the search to papers published between January 2015 and December 2020 that considered at least two echelons of the supply chain. In line with previous systematic reviews in the field, only journal papers, written in English and from an operations research perspective were included. Moreover, because the PM supply chains remain in place throughout, papers that analysed the healthcare emergency and humanitarian relief supply chains have not been considered. Finally, non-emergency healthcare supply chain papers with a focus on non-medical products were also excluded.

\begin{figure}[hbt!]
  \begin{subfigure}[b]{0.4\textwidth}
    \includegraphics[width=\textwidth]{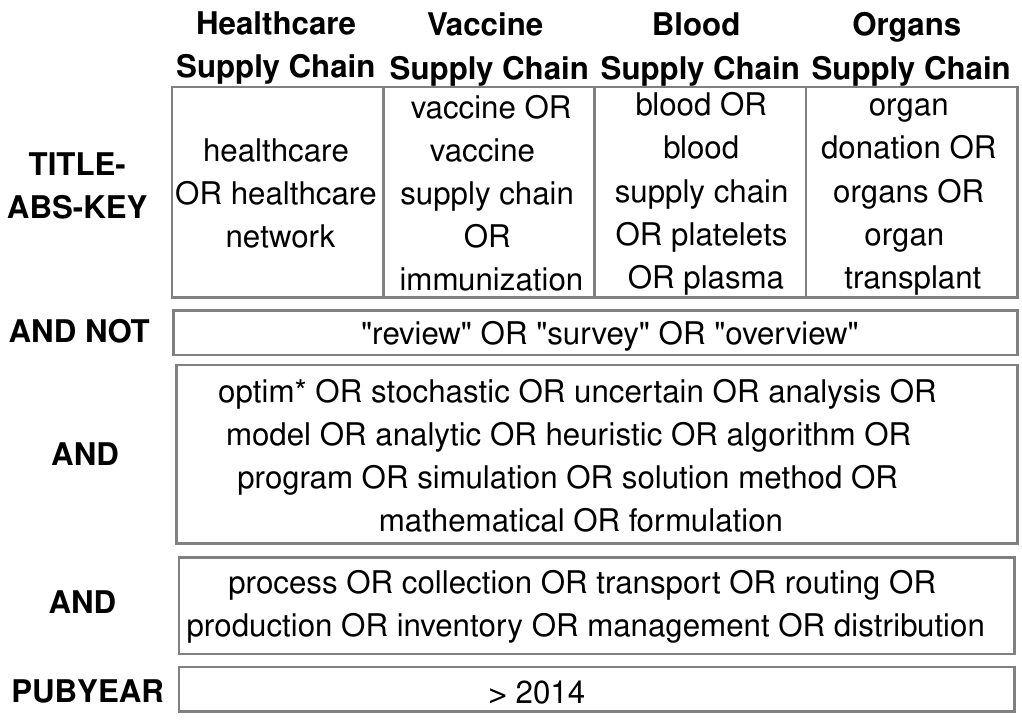}
    \caption[size = \footnotesize]{Search Strategy}
    \label{fig:search_strategy}
  \end{subfigure}
  \hfill
  \begin{subfigure}[b]{0.6\textwidth}
    \includegraphics[width=\textwidth]{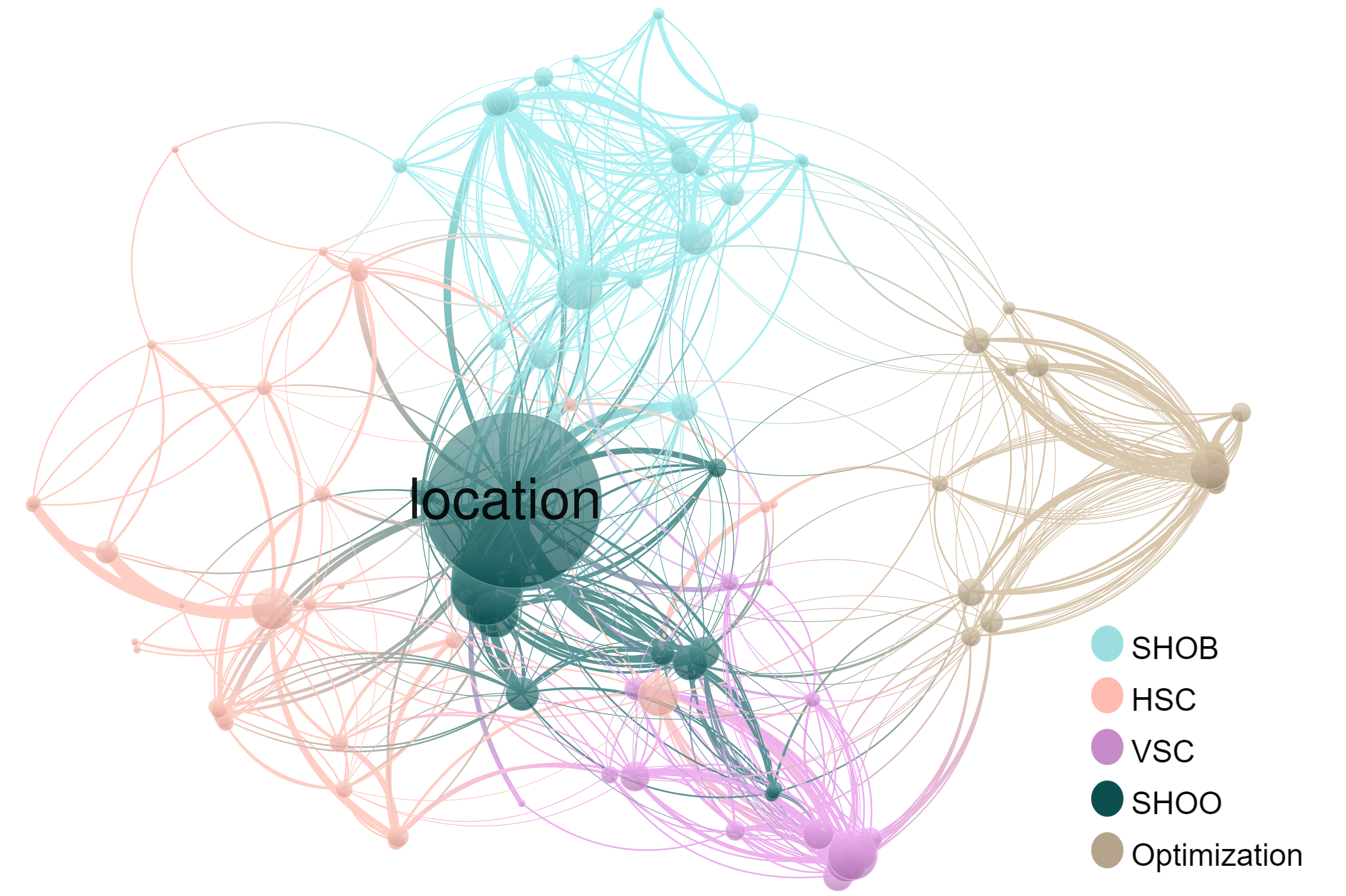}
    \caption{Topic Clustering}
    \label{fig:bibliographic_clustering}
  \end{subfigure}
  \caption{Keywords search (Figure \ref{fig:search_strategy}) and bibliometric analysis (Figure \ref{fig:bibliographic_clustering}) for the blood (SHOB), healthcare (HSC), vaccine (VSC), and organ donation (SHOO) supply chains.}
\end{figure}

The search strategy is presented in Figure \ref{fig:search_strategy}. The keywords were chosen starting from the supply chains of interest, namely non-emergency healthcare supply chain (including hospitals, nursing homes, and local treatment centres) (HSC), those that support substances of human origin (SHO), such as blood transfusions (SHOB) and organ transplantation (SHOO), and vaccines (VSC). The initial search returned 226 papers for SHOB, 45 papers for SHOO, and 44 papers for VSC, among which 202, 43, and 41 were excluded for each area respectively due to non-relevance. For HSC supply chain, the search returned over 1000 papers, but only 10 were included in the review either due to non-relevance or duplicates with the other searches. In addition, two conference papers and a journal article focusing on the optimisation of the CAR-T delivery, a class of autologous ATMP, were manually added to the bibliography. Finally, because the ATMPs are biopharmaceutical products, they share characteristics with the biopharmaceutical supply chain (BSC), and even though it is not extensively analysed in this review, references to it will be made in Section~\ref{subsec:integration_healthcare}.

An inspection of the literature was performed using text mining and network analysis on the title, keywords, and abstracts of the selected papers. The analysis returned five main clusters, which are colour coded in Figure~\ref{fig:bibliographic_clustering}. The size of each node represents the frequency of a keyword in all papers. The distance between nodes relates to how related the terms are between them. SHOB, HSC, VSC, and SHOO clusters differentiate between the commonly used terms specific to each of the four supply chains, while the optimization cluster belongs to modelling and solution methods. As expected, the keyword ``location'', forming the strategic level of these supply chains, is the most prevalent in the reviewed papers, indicating the high number of location problems in the field.

The complexity of the facility location problems (FLP) and their important role in the supply chain design made them receive particular interest and often be considered separately. Nonetheless, such an oversimplification can lead to finding multiple feasible or optimal solutions that work in isolation but fail to render similar results on an extended multi-echelon problem \cite{shen_qi_2007}. The increasing number of papers on integrated supply chain is also evident through the recent literature reviews, such as \cite{sharkey_et_al_2011} review on FLP with demand scheduling on predefined time windows; \cite{kaviani_2009} and \cite{farahani_et_al_2015} discussions on facility location and inventory management problems, and \cite{fahimnia_et_al_2013} review on production-distribution planning. In this paper, our focus is thus on proposing an interconnected model of the different echelons of the PM supply chain starting from the strategic level of FLP and extending it considering allocation and manufacturing optimization.

\section{Literature Review}

\subsection{Personalised Medicine Integration in Healthcare} \label{subsec:integration_healthcare}

\begin{figure}[hbt!]
    \centering
    \includegraphics{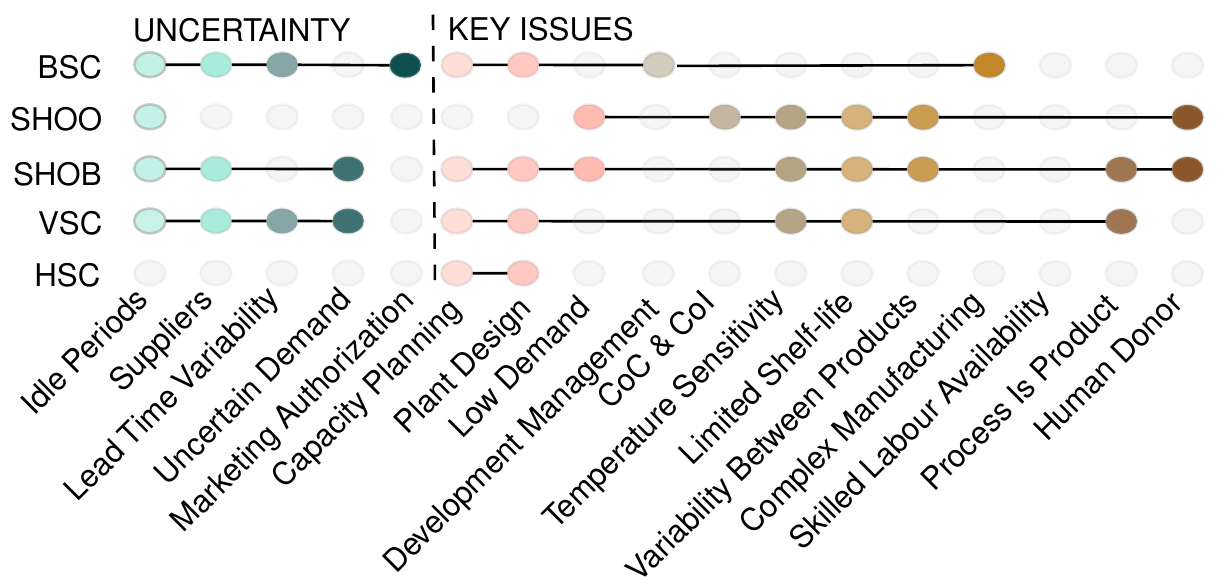}
    \caption{Comparative table showing the intersection of key issues and uncertainties between personalised medicine and other healthcare supply chains. CoC \& CoI refer to Chain of Custody and Chain of Identity respectively.}
    \label{fig:upset_plot}
\end{figure}

There is a considerable number of surveys of operation research methods applied to healthcare supply chains \citep[for reviews see, e.g.,][]{ahmadi-javid_et_al_2017b, duijzer_et_al_2018, piraban_et_al_2019}. The collection, storage, and transportation of medical products are mature and well-defined supply chains. An increased complexity is however brought by the ability of PM to create personalised re-engineered products, that are autologous in nature and force the medical treatment towards a patient-centred approach. The rest of this section discusses the uncertainties and key issues of the ATMP network design and how they intersect with those present in other healthcare supply chains. Figure~\ref{fig:upset_plot} gives an overview of this intersection.

The PM supply chain is prone to a number of uncertainties. More widely researched within the BSC, SHOs, and VSC supply chains, disruptions, idle periods, and supplier related uncertainties are concerns for the ATMPs. Lead time variability is characteristic to the VSC and BSC while the uncertain demand is challenging for the SHOB and VSC. Nonetheless, none of the latter two face a low demand that can lead to long periods of inactivity of parts of the supply chain. The autologous products typical of PM increase the difficulty of handling the uncertainties. While SHOs involve human donors, the process is completely allogeneic. The patient's health condition is not driving both the supply and demand. Moreover, in most cases, the SHOs products can be derived to other individuals, which is impossible for autologous products. 

The uncertain demand is strongly related to whether a product will obtain market authorization. The commercialisation approval is a lengthy process, dependent on the drug administration agencies. Together with the sensitive nature of their field, any changes in the manufacturing process of an ATMP after the initial commercial approval are highly regulated by authorities and require a lengthy reevaluation \citep{phillips_et_al_2011, iglesias-lopez_et_al_2019}. This makes it imperative for the supply chain to be optimised before market release.  
As a medical supply chain, PM shares common characteristics with other, more mature networks in the healthcare field. The first problem to be solved as part of the strategic level of the supply chain is a facility location problem. While the location of the hospitals is fixed, the location of manufacturing and cryopreservation facilities needs to be optimized. With the exception of SHOO, the other supply chains are also concerned with facility location and design problems. Nevertheless, the highly personalised products lead to low global demand for PM therapies and, alongside a long development process (BSC), makes the FLP more challenging. 

The PM has an agile supply chain driven by pull factors, meaning that the production is executed in response to a customer's needs. The usage of off-the-shelf products is no longer a viable approach in PM and the start and finish times of the supply chain are determined by the patient's health condition. Similar to the SHOO, there is thus a need for a Chain of Custody and Chain of Identity to be always maintained. Nevertheless, the high sensitive nature of genetic testing data that is usually associated with the ATMPs make issues such as privacy protection of even higher importance for PM \citep{miller_tucker_2018}.  Moreover, using living cells, the entire process has a short shelf-life. Therefore, similar to the SHOO, SHOB, and VSC, transportation is commonly done within a cold supply chain. However, in comparison to the more traditional stratified products, ATMPs add to the supply chain a \emph{complex manufacturing process} that is dictated by patient variability and the quality of the raw materials, which are characteristics of the biopharmaceutical supply chain (BSC).

With the increasing number of commercial ATMPs and the rapid growth of new technologies, the manufacturing process also suffers from a lack of skilled labour \citep{lewis_bradshaw_2017}. As a result, the shift towards a higher level of automation has received particular attention \citep{moutsatsou_et_al_2019}. As shown by \citet{lopes_et_al_2020}, this could reduce both the duration of the supply chain and the COGs. While the BSC also has manufacturing processes, the availability of skilled labours is not a major problem for any other healthcare supply chain. Additionally, each stage of the manufacturing process is influenced by various exogenous factors that can lead to lower quality therapies and ultimately become inefficient \citep{lipsitz_et_al_2016} hence, it is common to refer to the product as the entire process itself. Ensuring that the variability between products is minimal is one of the key aspects that need to be guaranteed should personalised medicine be widely implemented. 

\subsection{Objectives and Constraints in Healthcare Supply Chains} \label{sec:literature_review}

\begin{figure}[H]
\begin{minipage}[H]{.475\textwidth} 
    \includegraphics[width=\textwidth]{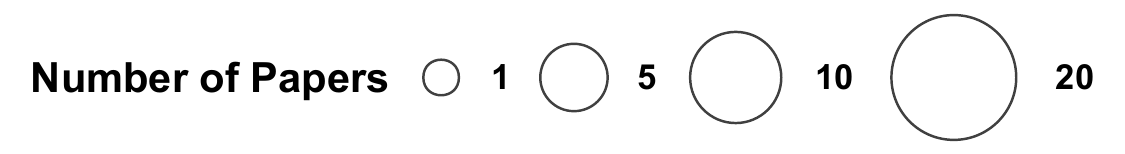}
\end{minipage}%
\hfill 
\begin{minipage}[H]{.475\textwidth}
    \includegraphics[width=\textwidth]{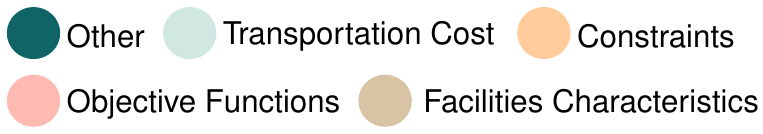}
\end{minipage}
    \centering
    \includegraphics[width=\textwidth]{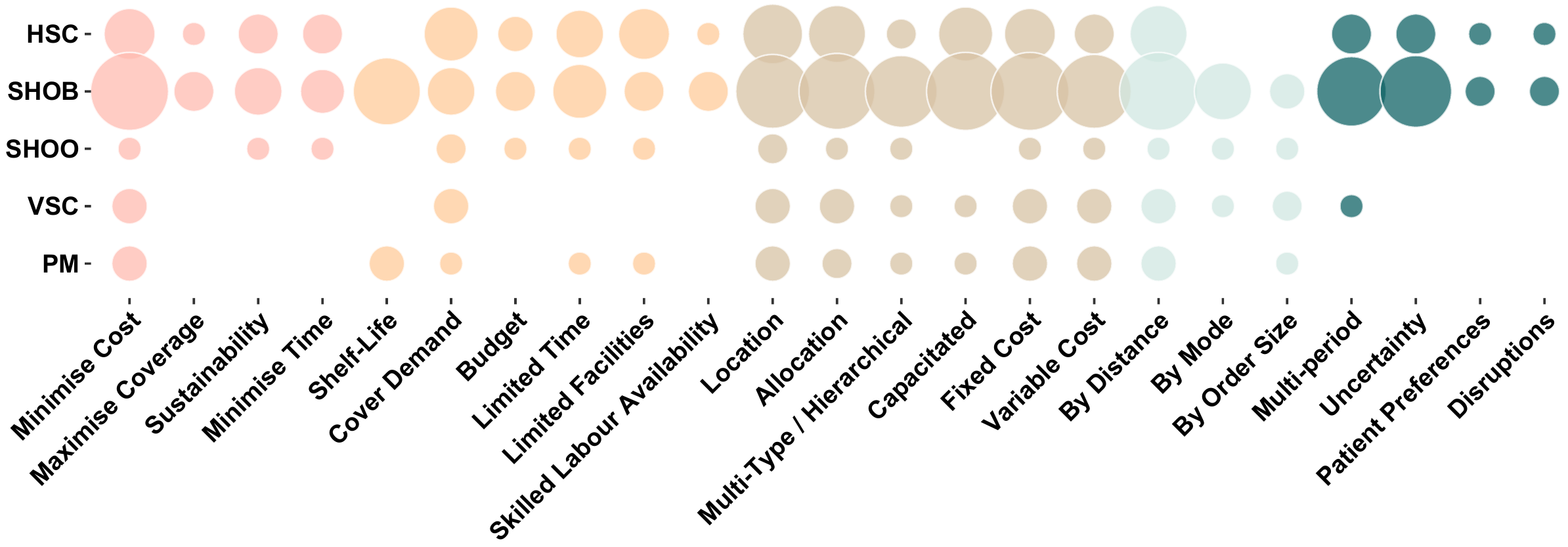}
    \caption{Bubble plot. \emph{x-axis}: personalised medicine characteristics, \emph{y-axis}: healthcare supply chains, \emph{size}: frequency in healthcare academic literature after 2015.}
    \label{bubble_plot}
\end{figure}

Following our discussion above of the characteristics of PM supply chains, we analyse now which objectives and constraints are mentioned in the reviewed papers as being relevant for each supply chain considered here. An overview of this analysis is shown in Figure \ref{bubble_plot}, where each bubble indicates, for each supply chain, how many papers considered a particular objective or constraint. A further breakdown of each reviewed paper is presented in Tables \ref{tab:literature_1} and \ref{tab:literature_2}.

Cost minimisation and profit maximisation were the most prevalent objectives across all four healthcare supply chains. In contrast to the public sector, which regulates most of the healthcare industry, the pharmaceutical sector is primarily driven by delivering drugs at the right time while maintaining the benefits for all stakeholders \citep{rossetti_et_al_2011}. Maximising coverage and ensuring social sustainability have not be extensively considered. Access to healthcare is an important ethical issue and, without significant improvements in the delivery methods of the ATMPs, personalised medicine will contribute to widening this barrier \citep{chong_et_al_2018}.  \citet{ares_et_al_2016}, \citet{eskandari-khanghahi_et_al_2018}, \citet{zhang_atkins_2019} and \citet{haeri_et_al_2020} are the only ones to consider social sustainability, either as equity and efficiency (usually defined by the waiting time or product shortages) or social welfare (usually defined by the impact of the supply chain on staff jobs). Similarly, environmental sustainability was discussed only by \citet{saif_elhedhli_2016}, \citet{heidari-fathian_pasandideh_2018}, and \citet{hamdan_diabat_2019}. They analysed this aspect either from a wastage perspective, CO2 emissions due to the cold supply chain, or the impacts of opening facilities. 

\begin{table}[hbt!]
\centering
\caption{The distribution of characteristics between the blood (SHOB), organ donation (SHOO), personalised medicine (PM), vaccine (VSC), and non-emergency healthcare (HSC) supply chains - Part 1.\label{tab:literature_1}}
\renewcommand{\arraystretch}{1}
\resizebox{\textwidth}{!}{%
\begin{tabular}{lllccccccccccccccc}
\toprule
     &                                         &  & \multicolumn{5}{c}{\textbf{Objective Functions}}                                                                                                                                            &                      & \multicolumn{6}{c}{\textbf{Constraints}}                                                                                                                                                                                                                                           &                      &                                         &                                              \\ \cline{4-7} \cline{10-15} \\
     
  \multicolumn{2}{l}{}                                                      
\begin{tikzpicture}[x=0.75pt,y=0.75pt,yscale=-1,xscale=1]

\draw   (220,20) -- (260.5,20) -- (260.5,30) -- (220,30) -- cycle ;
\draw   (220,40) -- (260.5,40) -- (260.5,50) -- (220,50) -- cycle ;
\draw   (220,61) -- (260.5,61) -- (260.5,71) -- (220,71) -- cycle ;
\draw   (220,81) -- (260.5,81) -- (260.5,91) -- (220,91) -- cycle ;
\draw   (220,100) -- (260.5,100) -- (260.5,110) -- (220,110) -- cycle ;
\draw  [color={gray}  ,draw opacity=1 ][fill={red}  ,fill opacity=1 ] (220,20) -- (260.5,20) -- (260.5,30) -- (220,30) -- cycle ;
\draw  [color={gray}  ,draw opacity=1 ][fill={purple}  ,fill opacity=1 ] (220,40) -- (260.5,40) -- (260.5,50) -- (220,50) -- cycle ;
\draw  [color={gray}  ,draw opacity=1 ][fill={ForestGreen}  ,fill opacity=1 ] (220,61) -- (260.5,61) -- (260.5,71) -- (220,71) -- cycle ;
\draw  [color={gray}  ,draw opacity=1 ][fill={blue}  ,fill opacity=1 ] (220,81) -- (260.5,81) -- (260.5,91) -- (220,91) -- cycle ;
\draw  [color={gray}  ,draw opacity=1 ][fill={yellow}  ,fill opacity=1 ] (220,100) -- (260.5,100) -- (260.5,110) -- (220,110) -- cycle ;
\draw  [color={gray}  ,draw opacity=1 ][fill={rgb, 255:red, 0; green, 0; blue, 0 }  ,fill opacity=0 ] (120,10) -- (178.5,10) -- (178.5,20) -- (120,20) -- cycle ;
\draw  [color={gray}  ,draw opacity=1 ][fill={rgb, 255:red, 0; green, 0; blue, 0 }  ,fill opacity=1 ] (120,10) -- (130.5,10) -- (130.5,20) -- (120,20) -- cycle ;
\draw  [color={gray}  ,draw opacity=1 ][fill={rgb, 255:red, 0; green, 0; blue, 0 }  ,fill opacity=1 ] (120,30) -- (140.5,30) -- (140.5,40) -- (120,40) -- cycle ;
\draw  [color={gray}  ,draw opacity=1 ][fill={rgb, 255:red, 0; green, 0; blue, 0 }  ,fill opacity=0 ] (120,30) -- (178.5,30) -- (178.5,40) -- (120,40) -- cycle ;
\draw  [color={gray}  ,draw opacity=1 ][fill={rgb, 255:red, 0; green, 0; blue, 0 }  ,fill opacity=0 ] (120,50) -- (178.5,50) -- (178.5,60) -- (120,60) -- cycle ;
\draw  [color={gray}  ,draw opacity=1 ][fill={rgb, 255:red, 0; green, 0; blue, 0 }  ,fill opacity=1 ] (120,50) -- (150.5,50) -- (150.5,60) -- (120,60) -- cycle ;
\draw  [color={gray}  ,draw opacity=1 ][fill={rgb, 255:red, 0; green, 0; blue, 0 }  ,fill opacity=1 ] (120,70) -- (160.5,70) -- (160.5,80) -- (120,80) -- cycle ;
\draw  [color={gray}  ,draw opacity=1 ][fill={rgb, 255:red, 0; green, 0; blue, 0 }  ,fill opacity=0 ] (120,70) -- (178.5,70) -- (178.5,80) -- (120,80) -- cycle ;
\draw  [color={gray}  ,draw opacity=1 ][fill={rgb, 255:red, 0; green, 0; blue, 0 }  ,fill opacity=0 ] (120,90) -- (178.5,90) -- (178.5,100) -- (120,100) -- cycle ;
\draw  [color={gray}  ,draw opacity=1 ][fill={rgb, 255:red, 0; green, 0; blue, 0 }  ,fill opacity=1 ] (120,90) -- (169.5,90) -- (169.5,100) -- (120,100) -- cycle ;
\draw  [color={gray}  ,draw opacity=1 ][fill={rgb, 255:red, 0; green, 0; blue, 0 }  ,fill opacity=1 ] (120,110) -- (178.5,110) -- (178.5,120) -- (120,120) -- cycle ;

\draw (268,59) node [anchor=north west][inner sep=0.75pt]  [font=\footnotesize] [align=left] {SHOB};
\draw (269,39) node [anchor=north west][inner sep=0.75pt]  [font=\footnotesize] [align=left] {PM};
\draw (269,79) node [anchor=north west][inner sep=0.75pt]  [font=\footnotesize] [align=left] {SHOO};
\draw (269,99) node [anchor=north west][inner sep=0.75pt]  [font=\footnotesize] [align=left] {VSC};
\draw (269,19) node [anchor=north west][inner sep=0.75pt]  [font=\footnotesize] [align=left] {HSC};
\draw (179.5,8) node [anchor=north west][inner sep=0.75pt]  [font=\footnotesize] [align=left] {2015};
\draw (179.5,28) node [anchor=north west][inner sep=0.75pt]  [font=\footnotesize] [align=left] {2016};
\draw (179.5,48) node [anchor=north west][inner sep=0.75pt]  [font=\footnotesize] [align=left] {2017};
\draw (179.5,68) node [anchor=north west][inner sep=0.75pt]  [font=\footnotesize] [align=left] {2018};
\draw (179.5,88) node [anchor=north west][inner sep=0.75pt]  [font=\footnotesize] [align=left] {2019};
\draw (179.5,108) node [anchor=north west][inner sep=0.75pt]  [font=\footnotesize] [align=left] {2020};

\end{tikzpicture} &  
	 & \rotatebox{90}{\rlap{\tabular{@{}l}\textbf{Minimise Cost} \endtabular}} 
	 & \rotatebox{90}{\rlap{\tabular{@{}l}\textbf{Maximise} \\ \textbf{Coverage} \endtabular}} 
	 & \rotatebox{90}{\rlap{\tabular{@{}l}\textbf{Social} \\ \textbf{Sustainability} \endtabular}} 
	 & \rotatebox{90}{\rlap{\tabular{@{}l}\textbf{Environmental} \\ \textbf{Sustainability} \endtabular}}  
     & \rotatebox{90}{\rlap{\tabular{@{}l}\textbf{Minimise Time} \endtabular}}
     & \multicolumn{1}{c}{} 
	 & \rotatebox{90}{\rlap{\tabular{@{}l}\textbf{Shelf-life} \endtabular}} 
     & \rotatebox{90}{\rlap{\tabular{@{}l}\textbf{Cover Demand} \endtabular}} 
     & \rotatebox{90}{\rlap{\tabular{@{}l}\textbf{Budget} \endtabular}} 
     & \rotatebox{90}{\rlap{\tabular{@{}l}\textbf{Limited Time} \endtabular}} 
     & \rotatebox{90}{\rlap{\tabular{@{}l}\textbf{Limited Facilities} \endtabular}} 
     & \rotatebox{90}{\rlap{\tabular{@{}l}\textbf{Skilled Labour} \\ \textbf{Availability} \endtabular}} 
     & \multicolumn{1}{c}{} 
	 & \rotatebox{90}{\rlap{\tabular{@{}l}\textbf{Case Study} \endtabular}}
     & \rotatebox{90}{\rlap{\tabular{@{}l}\textbf{Solution Method} \endtabular}} 
	 \\ 
	 \hline 
\drawBoxHSC{16}{10}{10}   & \citet{mestre_et_al_2015}               &  & \ding{109}       & \ding{109}           &              &             & \ding{109}        &    &              & \ding{109}      &                & \ding{109}    &                 &               &      & \ding{109}    & EM    \\
\drawBoxHSC{16}{10}{10}   & \citet{shishebori_babadi_2015}          &  & \ding{109}       &                      &              &             &                   &    &              &                 & \ding{109}     &               & \ding{109}      &               &      & \ding{109}    & EM    \\
\drawBoxHSC{16}{10}{10}   & \citet{cardoso_et_al_2015}              &  & \ding{109}       &                      & \ding{109}   &             &                   &    &              & \ding{109}      &                & \ding{109}    & \ding{109}      &               &      & \ding{109}    & EM    \\
\drawBoxHSC{32}{10}{10}   & \citet{ares_et_al_2016}                 &  &                  & \ding{109}           & \ding{109}   &             &                   &    &              &                 &                &               & \ding{109}      &               &      & \ding{109}    & IM     \\
\drawBoxHSC{48}{10}{10}   & \citet{zarrinpoor_et_al_2017}           &  & \ding{109}       &                      &              &             &                   &    &              & \ding{109}      &                & \ding{109}    & \ding{109}      &               &      & \ding{109}    & EM    \\
\drawBoxHSC{64}{10}{10}   & \citet{wang_ma_2018}                    &  & \ding{109}       &                      &              &             & \ding{109}        &    &              & \ding{109}      &                &               & \ding{109}      &               &      & \ding{109}    & IM    \\
\drawBoxHSC{80}{10}{10}   & \citet{vieira_et_al_2019}               &  &                  &                      &              &             & \ding{109}        &    &              & \ding{109}      &                & \ding{109}    & \ding{109}      &               &      & \ding{109}    & EM    \\
\drawBoxHSC{80}{10}{10}   & \citet{zhang_atkins_2019}               &  &                  &                      & \ding{109}   &             &                   &    &              &                 &                &               & \ding{109}      &               &      & \ding{109}    & IM    \\
\drawBoxHSC{80}{10}{10}   & \citet{dogan_et_al_2019}                &  & \ding{109}       &                      &              &             & \ding{109}        &    &              & \ding{109}      & \ding{109}     & \ding{109}    &                 &               &      & \ding{109}    & EM    \\
\drawBoxHSC{100}{10}{10}  & \citet{mendoza-gomez_et_al_2020}        &  & \ding{109}       &                      &              &             &                   &    &              & \ding{109}      &                &               &                 &               &      &               & IM    \\
\drawBoxPM{64}{10}{10}    & \citet{wang_et_al_2018}                 &  & \ding{109}       &                      &              &             &                   &    & \ding{109}   &                 &                &               &                 &               &      &               & EM    \\
\drawBoxPM{100}{10}{10}   & \citet{moschou_et_al_2020}              &  & \ding{109}       &                      &              &             &                   &    & \ding{109}   &                 &                &               &                 &               &      & \ding{109}    & EM    \\
\drawBoxPM{100}{10}{10}   & \citet{karakostas_et_al_2020}           &  & \ding{109}       &                      &              &             &                   &    & \ding{109}   & \ding{109}      &                & \ding{109}    &                 &               &      &               & IM    \\
\drawBoxSHOB{16}{10}{10}  & \citet{zahiri_et_al_2015}               &  & \ding{109}       &                      &              &             &                   &    & \ding{109}   & \ding{109}      & \ding{109}     & \ding{109}    & \ding{109}      &     &      & \ding{109}    & EM    \\
\drawBoxSHOB{16}{10}{10}  & \citet{elalouf_et_al_2015}              &  & \ding{109}       &                      &              &             &                   &    &              &                 &                &               &                 &               &      & \ding{109}    & EM    \\
\drawBoxSHOB{16}{10}{10}  & \citet{arvan_et_al_2015}                &  & \ding{109}       &                      &              &             &                   &    & \ding{109}   & \ding{109}      & \ding{109}     &               &                 &               &      &               & EM    \\
\drawBoxSHOB{32}{10}{10}  & \citet{chaiwuttisak_et_al_2016}         &  & \ding{109}       &                      &              &             & \ding{109}        &    & \ding{109}   &                 &                & \ding{109}    &                 &               &      & \ding{109}    & EM    \\
\drawBoxSHOB{48}{10}{10}  & \citet{fahimnia_et_al_2017}             &  & \ding{109}       &                      &              &             & \ding{109}        &    &              & \ding{109}      &                & \ding{109}    & \ding{109}      &     &      &               & EM    \\
\drawBoxSHOB{48}{10}{10}  & \citet{osorio_et_al_2017}               &  & \ding{109}       &                      &              &             &                   &    & \ding{109}   & \ding{109}      &                &               &                 &               &      & \ding{109}    & EM    \\
\drawBoxSHOB{48}{10}{10}  & \citet{ramezanian_behboodi_2017}        &  & \ding{109}       & \ding{109}           &              & \ding{109}  &                   &    & \ding{109}   &                 &                &               &                 &               &      & \ding{109}    & EM    \\
\drawBoxSHOB{48}{10}{10}  & \citet{ensafian_yaghoubi_2017}          &  &                  & \ding{109}           &              &             & \ding{109}        &    & \ding{109}   &                 &                &               &                 &               &      &  \ding{109}     & EM      \\
\drawBoxSHOB{48}{10}{10}  & \citet{zahiri_pishvaee_2017}            &  & \ding{109}       &                      &              &             &                   &    & \ding{109}   &                 & \ding{109}     &               & \ding{109}      &               &      & \ding{109}    & EM    \\
\drawBoxSHOB{48}{10}{10}  & \citet{ensafian_et_al_2017}             &  & \ding{109}       &                      &              &             &                   &    & \ding{109}   &                 &                &               &                 &               &      & \ding{109}    & EM    \\
\drawBoxSHOB{48}{10}{10}  & \citet{attari_et_al_2018}               &  & \ding{109}       &                      &              &             & \ding{109}        &    & \ding{109}   &                 &                & \ding{109}    &                 &               &      & \ding{109}    & EM    \\
\drawBoxSHOB{64}{10}{10}  & \citet{eskandari-khanghahi_et_al_2018}  &  & \ding{109}       &                      & \ding{109}   & \ding{109}  &                   &    & \ding{109}   &                 &                &               &                 &               &      &               & IM    \\
\drawBoxSHOB{64}{10}{10}  & \citet{heidari-fathian_pasandideh_2018} &  & \ding{109}       &                      & \ding{109}   & \ding{109}  &                   &    & \ding{109}   &                 &                &               &                 &               &      &               & EM    \\
\drawBoxSHOB{64}{10}{10}  & \citet{osorio_et_al_2018}               &  & \ding{109}       &                      &              &             &                   &    &              & \ding{109}      &                & \ding{109}    &                 &               &      & \ding{109}    &  EM     \\
\drawBoxSHOB{64}{10}{10}  & \citet{samani_hosseini-motlagh_2018}    &  & \ding{109}       &                      &              &             &                   &    & \ding{109}   &                 &                & \ding{109}    &                 &               &      & \ding{109}    & EM    \\
\drawBoxSHOB{80}{10}{10}  & \citet{bruno_et_al_2019}                &  & \ding{109}       &                      &              &             &                   &    &              &                 &                & \ding{109}    &                 &               &      & \ding{109}    & EM    \\
\drawBoxSHOB{80}{10}{10}  & \citet{hamdan_diabat_2019}              &  & \ding{109}       &                      &              & \ding{109}  & \ding{109}        &    & \ding{109}   & \ding{109}      &                &               & \ding{109}      &               &      & \ding{109}    & EM    \\
\drawBoxSHOB{100}{10}{10} & \citet{haeri_et_al_2020}                &  & \ding{109}       & \ding{109}           & \ding{109}   &             &                   &    &              &                 &                & \ding{109}    &                 &     &      & \ding{109}    & EM    \\
\drawBoxSHOB{100}{10}{10} & \citet{hosseini-motlagh_et_al_2020}     &  & \ding{109}       &                      &              &             &                   &    &              &                 & \ding{109}     & \ding{109}    &                 & \ding{109}    &      & \ding{109}    & EM    \\
\drawBoxSHOB{100}{10}{10} & \citet{hosseini-motlagh_et_al_2020b} &	& \ding{109}	&	&	& & &	& \ding{109}	& \ding{109}	&	& \ding{109}	& \ding{109}	& &	& \ding{109}	& EM \\
\drawBoxSHOO{48}{10}{10}  & \citet{rajmohan_et_al_2017}             &  &                  &                      &              &             & \ding{109}        &    &              & \ding{109}      &                &               & \ding{109}      &               &      & \ding{109}    & EM    \\
\drawBoxSHOO{80}{10}{10}  & \citet{rabbani_talebi_2019}             &  & \ding{109}       &                      & \ding{109}   &             &                   &    &              & \ding{109}      & \ding{109}     & \ding{109}    &                 &               &      & \ding{109}    & EM    \\
\drawBoxVSC{32}{10}{10}   & \citet{saif_elhedhli_2016}              &  & \ding{109}       &                      &              & \ding{109}  &                   &    &              & \ding{109}      &                &               &                 &               &      & \ding{109}    & EM    \\
\drawBoxVSC{80}{10}{10}   & \citet{lim_et_al_2019}                  &  & \ding{109}       &                      &              &             &                   &    &              & \ding{109}      &                &               &                 &               &      & \ding{109}    & EM    \\
\drawBoxVSC{100}{10}{10}  & \citet{yang_et_al_2020}                 &  & \ding{109}       &                      &              &             &                   &    &              & \ding{109}      &                &               &                 &               &      & \ding{109}    & EM    \\ 
\bottomrule
\multicolumn{17}{l}{\textbf{Note:} \textit{Social Sustainability} includes efficiency and equity. For \textit{Solution Method} EM is Exact Methods and IM is Inexact Methods.}
\end{tabular}%
}
\end{table}

\begin{table}[hbt!]
\centering
\caption{The distribution of characteristics between the blood (SHOB), organ donation (SHOO), personalised medicine (PM), vaccine (VSC), and non-emergency healthcare (HSC) supply chains - Part 2.\label{tab:literature_2}}
\renewcommand{\arraystretch}{1}
\resizebox{\textwidth}{!}{%
\begin{tabular}{@{}lllccccccccccccccc@{}}
\toprule
     &                                    &  & \multicolumn{6}{c}{\textbf{Facilities Characteristics}}                                                            & \textbf{} & \multicolumn{3}{c}{\textbf{Delivery Cost}}  & \textbf{} & \multicolumn{4}{c}{\textbf{Other}}                                    \\ 
	 \cmidrule(lr){4-9} \cmidrule(lr){11-13} \cmidrule(l){15-18} 
     & \begin{tikzpicture}[x=0.75pt,y=0.75pt,yscale=-1,xscale=1]

\draw   (220,20) -- (260.5,20) -- (260.5,30) -- (220,30) -- cycle ;
\draw   (220,40) -- (260.5,40) -- (260.5,50) -- (220,50) -- cycle ;
\draw   (220,61) -- (260.5,61) -- (260.5,71) -- (220,71) -- cycle ;
\draw   (220,81) -- (260.5,81) -- (260.5,91) -- (220,91) -- cycle ;
\draw   (220,100) -- (260.5,100) -- (260.5,110) -- (220,110) -- cycle ;
\draw  [color={gray}  ,draw opacity=1 ][fill={red}  ,fill opacity=1 ] (220,20) -- (260.5,20) -- (260.5,30) -- (220,30) -- cycle ;
\draw  [color={gray}  ,draw opacity=1 ][fill={purple}  ,fill opacity=1 ] (220,40) -- (260.5,40) -- (260.5,50) -- (220,50) -- cycle ;
\draw  [color={gray}  ,draw opacity=1 ][fill={ForestGreen}  ,fill opacity=1 ] (220,61) -- (260.5,61) -- (260.5,71) -- (220,71) -- cycle ;
\draw  [color={gray}  ,draw opacity=1 ][fill={blue}  ,fill opacity=1 ] (220,81) -- (260.5,81) -- (260.5,91) -- (220,91) -- cycle ;
\draw  [color={gray}  ,draw opacity=1 ][fill={yellow}  ,fill opacity=1 ] (220,100) -- (260.5,100) -- (260.5,110) -- (220,110) -- cycle ;
\draw  [color={gray}  ,draw opacity=1 ][fill={rgb, 255:red, 0; green, 0; blue, 0 }  ,fill opacity=0 ] (120,10) -- (178.5,10) -- (178.5,20) -- (120,20) -- cycle ;
\draw  [color={gray}  ,draw opacity=1 ][fill={rgb, 255:red, 0; green, 0; blue, 0 }  ,fill opacity=1 ] (120,10) -- (130.5,10) -- (130.5,20) -- (120,20) -- cycle ;
\draw  [color={gray}  ,draw opacity=1 ][fill={rgb, 255:red, 0; green, 0; blue, 0 }  ,fill opacity=1 ] (120,30) -- (140.5,30) -- (140.5,40) -- (120,40) -- cycle ;
\draw  [color={gray}  ,draw opacity=1 ][fill={rgb, 255:red, 0; green, 0; blue, 0 }  ,fill opacity=0 ] (120,30) -- (178.5,30) -- (178.5,40) -- (120,40) -- cycle ;
\draw  [color={gray}  ,draw opacity=1 ][fill={rgb, 255:red, 0; green, 0; blue, 0 }  ,fill opacity=0 ] (120,50) -- (178.5,50) -- (178.5,60) -- (120,60) -- cycle ;
\draw  [color={gray}  ,draw opacity=1 ][fill={rgb, 255:red, 0; green, 0; blue, 0 }  ,fill opacity=1 ] (120,50) -- (150.5,50) -- (150.5,60) -- (120,60) -- cycle ;
\draw  [color={gray}  ,draw opacity=1 ][fill={rgb, 255:red, 0; green, 0; blue, 0 }  ,fill opacity=1 ] (120,70) -- (160.5,70) -- (160.5,80) -- (120,80) -- cycle ;
\draw  [color={gray}  ,draw opacity=1 ][fill={rgb, 255:red, 0; green, 0; blue, 0 }  ,fill opacity=0 ] (120,70) -- (178.5,70) -- (178.5,80) -- (120,80) -- cycle ;
\draw  [color={gray}  ,draw opacity=1 ][fill={rgb, 255:red, 0; green, 0; blue, 0 }  ,fill opacity=0 ] (120,90) -- (178.5,90) -- (178.5,100) -- (120,100) -- cycle ;
\draw  [color={gray}  ,draw opacity=1 ][fill={rgb, 255:red, 0; green, 0; blue, 0 }  ,fill opacity=1 ] (120,90) -- (169.5,90) -- (169.5,100) -- (120,100) -- cycle ;
\draw  [color={gray}  ,draw opacity=1 ][fill={rgb, 255:red, 0; green, 0; blue, 0 }  ,fill opacity=1 ] (120,110) -- (178.5,110) -- (178.5,120) -- (120,120) -- cycle ;

\draw (268,59) node [anchor=north west][inner sep=0.75pt]  [font=\footnotesize] [align=left] {SHOB};
\draw (269,39) node [anchor=north west][inner sep=0.75pt]  [font=\footnotesize] [align=left] {PM};
\draw (269,79) node [anchor=north west][inner sep=0.75pt]  [font=\footnotesize] [align=left] {SHOO};
\draw (269,99) node [anchor=north west][inner sep=0.75pt]  [font=\footnotesize] [align=left] {VSC};
\draw (269,19) node [anchor=north west][inner sep=0.75pt]  [font=\footnotesize] [align=left] {HSC};
\draw (179.5,8) node [anchor=north west][inner sep=0.75pt]  [font=\footnotesize] [align=left] {2015};
\draw (179.5,28) node [anchor=north west][inner sep=0.75pt]  [font=\footnotesize] [align=left] {2016};
\draw (179.5,48) node [anchor=north west][inner sep=0.75pt]  [font=\footnotesize] [align=left] {2017};
\draw (179.5,68) node [anchor=north west][inner sep=0.75pt]  [font=\footnotesize] [align=left] {2018};
\draw (179.5,88) node [anchor=north west][inner sep=0.75pt]  [font=\footnotesize] [align=left] {2019};
\draw (179.5,108) node [anchor=north west][inner sep=0.75pt]  [font=\footnotesize] [align=left] {2020};

\end{tikzpicture}
     &
	 & \rotatebox{90}{\rlap{\tabular{@{}l}\textbf{Location} \endtabular}} 
	 & \rotatebox{90}{\rlap{\tabular{@{}l}\textbf{Allocation} \endtabular}} 
	 & \rotatebox{90}{\rlap{\tabular{@{}l}\textbf{Multi-Type /} \\ \textbf{Hierarchical} \endtabular}} 
     & \rotatebox{90}{\rlap{\tabular{@{}l}\textbf{Capacitated} \endtabular}}
	 & \rotatebox{90}{\rlap{\tabular{@{}l}\textbf{Fixed Cost} \endtabular}}
	 & \rotatebox{90}{\rlap{\tabular{@{}l}\textbf{Variable Cost} \endtabular}}
     & \multicolumn{1}{c}{}
	 & \rotatebox{90}{\rlap{\tabular{@{}l}\textbf{By Distance} \endtabular}}
	 & \rotatebox{90}{\rlap{\tabular{@{}l}\textbf{By Mode} \endtabular}}
	 & \rotatebox{90}{\rlap{\tabular{@{}l}\textbf{By Order Size} \endtabular}}
	 & \multicolumn{1}{c}{}
	 & \rotatebox{90}{\rlap{\tabular{@{}l}\textbf{Multi-period} \endtabular}}
	 & \rotatebox{90}{\rlap{\tabular{@{}l}\textbf{Uncertainty} \endtabular}}
	 & \rotatebox{90}{\rlap{\tabular{@{}l}\textbf{Patient} \\ \textbf{Preferences} \endtabular}}
	 & \rotatebox{90}{\rlap{\tabular{@{}l}\textbf{Disruptions} \endtabular}}	 
	 \\ 
	 \hline
\drawBoxHSC{16}{10}{10}   & \citet{mestre_et_al_2015}               &  & \ding{109}          & \ding{109}            &                          & \ding{109}  & \ding{109}  & \ding{109}    &           & \ding{109}    &              &                    &           & \ding{109}   & \ding{109}  &                            & \ding{109}  \\
\drawBoxHSC{16}{10}{10}   & \citet{shishebori_babadi_2015}          &  & \ding{109}          &                       &                          & \ding{109}  &             &               &           & \ding{109}    &              &                    &           &              &             &                            &             \\
\drawBoxHSC{16}{10}{10}   & \citet{cardoso_et_al_2015}              &  & \ding{109}          & \ding{109}            &                          & \ding{109}  & \ding{109}  & \ding{109}    &           & \ding{109}    &              &                    &           &              &             &                            &             \\
\drawBoxHSC{32}{10}{10}   & \citet{ares_et_al_2016}                 &  & \ding{109}          &                       &                          &             &             &               &           & \ding{109}    &              &                    &           &              & \ding{109}  &                            &             \\
\drawBoxHSC{48}{10}{10}   & \citet{zarrinpoor_et_al_2017}           &  & \ding{109}          & \ding{109}            & \ding{109}               &             & \ding{109}  &               &           & \ding{109}    &              &                    &           &              & \ding{109}  & \ding{109}                 &             \\
\drawBoxHSC{64}{10}{10}   & \citet{wang_ma_2018}                    &  & \ding{109}          & \ding{109}            & \ding{109}               & \ding{109}  & \ding{109}  &               &           &               &              &                    &           &              &             &                            &             \\
\drawBoxHSC{80}{10}{10}   & \citet{vieira_et_al_2019}               &  & \ding{109}          & \ding{109}            &                          &             &             &               &           & \ding{109}    &              &                    &           &              &             &                            &             \\
\drawBoxHSC{80}{10}{10}   & \citet{zhang_atkins_2019}               &  & \ding{109}          & \ding{109}            &                          & \ding{109}  &             &               &           & \ding{109}    &              &                    &           & \ding{109}   &             & \ding{109}                 &             \\
\drawBoxHSC{80}{10}{10}   & \citet{dogan_et_al_2019}                &  & \ding{109}          & \ding{109}            &                          & \ding{109}  & \ding{109}  &               &           & \ding{109}    &              &                    &           &              &             &                            &             \\
\drawBoxHSC{100}{10}{10}  & \citet{mendoza-gomez_et_al_2020}        &  &                     & \ding{109}            &                          & \ding{109}  & \ding{109}  & \ding{109}    &           & \ding{109}    &              &                    &           & \ding{109}   & \ding{109}  &                            &             \\
\drawBoxPM{64}{10}{10}    & \citet{wang_et_al_2018}                 &  & \ding{109}          &                       &                          &             & \ding{109} & \ding{109}   &           & \ding{109}   &              & \ding{109}        &           &              &             &                            &             \\
\drawBoxPM{100}{10}{10}   & \citet{moschou_et_al_2020}              &  & \ding{109}          & \ding{109}            &                          &             & \ding{109}  & \ding{109}    &           & \ding{109}    &              &                    &           &              &             &                            &             \\
\drawBoxPM{100}{10}{10}   & \citet{karakostas_et_al_2020}           &  & \ding{109}          & \ding{109}            & \ding{109}               & \ding{109}  & \ding{109}  & \ding{109}    &           & \ding{109}    &              &                    &           &              &             &                            &             \\
\drawBoxSHOB{16}{10}{10}  & \citet{zahiri_et_al_2015}               &  & \ding{109}          & \ding{109}            & \ding{109}               & \ding{109}  & \ding{109}  &               &           & \ding{109}    &              & \ding{109}         &           &              & \ding{109}  &                            &             \\
\drawBoxSHOB{16}{10}{10}  & \citet{elalouf_et_al_2015}              &  & \ding{109}          & \ding{109}            & \ding{109}               &             & \ding{109}  &               &           & \ding{109}    &              &                    &           & \ding{109}   & \ding{109}  &                            &             \\
\drawBoxSHOB{16}{10}{10}  & \citet{arvan_et_al_2015}                &  & \ding{109}          & \ding{109}            & \ding{109}               & \ding{109}  & \ding{109}  & \ding{109}    &           & \ding{109}    &              &                    &           &              &             &                            &             \\
\drawBoxSHOB{32}{10}{10}  & \citet{chaiwuttisak_et_al_2016}         &  & \ding{109}          &                       & \ding{109}               & \ding{109}  & \ding{109}  & \ding{109}    &           & \ding{109}    & \ding{109}   &                    &           &              &             &                            &             \\
\drawBoxSHOB{48}{10}{10}  & \citet{fahimnia_et_al_2017}             &  & \ding{109}          & \ding{109}            &                          & \ding{109}  & \ding{109}  &               &           & \ding{109}    &              &                    &           & \ding{109}   &             &                            &             \\
\drawBoxSHOB{48}{10}{10}  & \citet{osorio_et_al_2017}               &  & \ding{109}          & \ding{109}            &                          & \ding{109}  &             & \ding{109}    &           &               &              &                    &           &              & \ding{109}  &                            &             \\
\drawBoxSHOB{48}{10}{10}  & \citet{ramezanian_behboodi_2017}        &  & \ding{109}          & \ding{109}            & \ding{109}               & \ding{109}  & \ding{109}  & \ding{109}    &           & \ding{109}    & \ding{109}   &                    &           & \ding{109}   & \ding{109}  &                            &             \\
\drawBoxSHOB{48}{10}{10}  & \citet{ensafian_yaghoubi_2017}          &  &                     & \ding{109}            & \ding{109}               & \ding{109}  & \ding{109}  & \ding{109}    &           & \ding{109}    & \ding{109}   &                    &           & \ding{109}   & \ding{109}  &                            &             \\
\drawBoxSHOB{48}{10}{10}  & \citet{zahiri_pishvaee_2017}            &  &                     & \ding{109}            & \ding{109}               & \ding{109}  & \ding{109}  & \ding{109}   &           & \ding{109}    & \ding{109}   &                    &           & \ding{109}   & \ding{109}  &                            &             \\
\drawBoxSHOB{48}{10}{10}  & \citet{ensafian_et_al_2017}             &  &                     & \ding{109}            &                          & \ding{109}  & \ding{109}  & \ding{109}    &           & \ding{109}    & \ding{109}   &                    &           & \ding{109}   & \ding{109}  &                            &             \\
\drawBoxSHOB{48}{10}{10}  & \citet{attari_et_al_2018}               &  & \ding{109}          & \ding{109}            &                          & \ding{109}  & \ding{109}  & \ding{109}    &           & \ding{109}    &              &                    &           & \ding{109}   & \ding{109}  &                            &             \\
\drawBoxSHOB{64}{10}{10}  & \citet{eskandari-khanghahi_et_al_2018}  &  & \ding{109}          & \ding{109}            & \ding{109}               & \ding{109}  & \ding{109}  & \ding{109}    &           & \ding{109}    & \ding{109}   & \ding{109}         &           & \ding{109}   & \ding{109}  &                            &             \\
\drawBoxSHOB{64}{10}{10}  & \citet{heidari-fathian_pasandideh_2018} &  & \ding{109}          & \ding{109}            & \ding{109}               & \ding{109}  & \ding{109}  & \ding{109}    &           & \ding{109}    & \ding{109}   & \ding{109}         &           & \ding{109}   & \ding{109}  &                            &             \\
\drawBoxSHOB{64}{10}{10}  & \citet{osorio_et_al_2018}               &  & \ding{109}          &                       & \ding{109}               & \ding{109}  & \ding{109}  & \ding{109}    &           & \ding{109}    &              &                    &           &              & \ding{109}  &                            &             \\
\drawBoxSHOB{64}{10}{10}  & \citet{samani_hosseini-motlagh_2018}    &  & \ding{109}          & \ding{109}            & \ding{109}               & \ding{109}  & \ding{109}  & \ding{109}    &           & \ding{109}    &              &                    &           & \ding{109}   & \ding{109}  &                            &             \\
\drawBoxSHOB{80}{10}{10}  & \citet{bruno_et_al_2019}                &  & \ding{109}          & \ding{109}            & \ding{109}               & \ding{109}  & \ding{109}  & \ding{109}    &           & \ding{109}    &              &                    &           &              &             &                            &             \\
\drawBoxSHOB{80}{10}{10}  & \citet{hamdan_diabat_2019}              &  & \ding{109}          & \ding{109}            & \ding{109}               & \ding{109}  & \ding{109}  & \ding{109}    &           & \ding{109}    & \ding{109}   &                    &           & \ding{109}   & \ding{109}  &                            &             \\
\drawBoxSHOB{100}{10}{10}  & \citet{haeri_et_al_2020}                &  & \ding{109}          & \ding{109}            & \ding{109}               & \ding{109}  & \ding{109}  & \ding{109}    &           & \ding{109}    &              &                    &           & \ding{109}   & \ding{109}  & \ding{109}                 & \ding{109}  \\
\drawBoxSHOB{100}{10}{10}  & \citet{hosseini-motlagh_et_al_2020}     &  & \ding{109}          & \ding{109}            & \ding{109}               & \ding{109}  & \ding{109}  & \ding{109}    &           & \ding{109}    & \ding{109}   &                    &           & \ding{109}   &   &                            & \ding{109}  \\
\drawBoxSHOB{100}{10}{10} & \citet{hosseini-motlagh_et_al_2020b} & & \ding{109}	& \ding{109}	& \ding{109}	& \ding{109}	&	& \ding{109}	& \ding{109}	&	&	& \ding{109}	& \ding{109}	& \ding{109}	& \ding{109} \\
\drawBoxSHOO{48}{10}{10}  & \citet{rajmohan_et_al_2017}             &  & \ding{109}          &                       &                          &             &             &               &           &               &              &                    &           &              &             &                            &             \\
\drawBoxSHOO{80}{10}{10}  & \citet{rabbani_talebi_2019}             &  & \ding{109}          & \ding{109}            & \ding{109}               &             & \ding{109}  & \ding{109}    &           & \ding{109}    & \ding{109}   & \ding{109}         &           & \ding{109}   &             &                            &             \\
\drawBoxVSC{32}{10}{10}   & \citet{saif_elhedhli_2016}              &  & \ding{109}          & \ding{109}            &                          &             & \ding{109}  & \ding{109}    &           & \ding{109}    & \ding{109}   &                    &           &              &             &                            &             \\
\drawBoxVSC{80}{10}{10}   & \citet{lim_et_al_2019}                  &  & \ding{109}          & \ding{109}            & \ding{109}               &             & \ding{109}  & \ding{109}    &           & \ding{109}    &              & \ding{109}         &           & \ding{109}   &             &                            &             \\
\drawBoxVSC{100}{10}{10}  & \citet{yang_et_al_2020}                 &  & \ding{109}          & \ding{109}            &                          & \ding{109}  & \ding{109}  & \ding{109}    &           & \ding{109}    &              & \ding{109}         &           &              &             &                            &             \\ 
\bottomrule
\end{tabular}%
}
\end{table}


\subsection{Gaps in the Literature of Personalised Medicine Supply Chain}

Generally, the non-emergency supply chains are designed considering a stable environment. External disruptions (e.g. lack of demand) are not of main concern. Regardless, the reviewed studies do not give sufficient consideration to internal disruptions, such as workers unavailability, equipment breakdown, or changes in the scheduling process. With potentially disastrous consequences for the patient in case of any interruptions, the supply chain resilience in terms of both resistance and recovery becomes highly relevant for ATMPs. Finally, transportation (modelled as a function of distance and, to a lesser extent, order size) has been largely considered using only one type of vehicle. It seems possible that this simplification is a result of the limited geographical areas considered in the papers. Most of the case studies are restricted to local or national areas and using a relatively sparse granularity. 

As also emphasised in Section \ref{sec:integration_healthcare}, availability of skilled labour is the least prevalent constraint in the reviewed papers. \cite{hosseini-motlagh_et_al_2020} are the only ones to consider skilled labours as part of their problem by directly linking the staff's proficiency with the patient's satisfaction. No reviewed paper was concerned with a lack of labour resources. Similarly, one of the biggest discrepancies between ATMPs and other supply chains, with few exceptions, is the lack of consideration for patient or user preferences. This is a direct consequence of the exclusive allogeneic (material collected from a donor) nature of the other supply chains, where an off-the-shelf approach with inventory management is implemented. If PM is to replace the traditional pharmaceutical products, the logistics of the treatment will ultimately be the patient's choice. Understanding patient preferences \cite{liu_et_al_2018} and ensuring fairness \cite{qi_2017} has been considered in the past in the healthcare management and operations research literature but has not been directly considered in relation to PM optimisation.  

In the context of PM, \citet{wang_et_al_2018, moschou_et_al_2020} and \citet{karakostas_et_al_2020} were the only ones that discussed the supply chain from an operations research perspective. The three papers look at the supply chain of CAR-T, a type of autologous ATMP. Nevertheless, none of the proposed models is exhaustive. The papers do not consider patient preferences, disruptions, resources availability or different risk levels as part of their models. With the exception of \citet{wang_et_al_2018}, who also aims to minimise the response time of the supply chain, the only objective is to minimise the cost of the network. Finding trade-offs between hospitals coverage and the equity and efficiency of the supply chain alongside time and cost minimisation could potentially lead to more realistic models of PM supply chain. 

\newpage

\section{Healthcare Mathematical Models} \label{sec:math_models}

The extensive number of papers and models leads to a clear distinction in the literature between the research focused on facility-location problems (FLP) and the one considering the wider umbrella of SCM. In this context, \cite{melo_et_al_2009} reviewed the discrete FLP literature of supply chain network, focusing on papers integrating capacity, inventory, procurement, production, routing, and transportation alongside financial aspects and risk management. Other reviews of integrated models of supply chain network design (SCND) were conducted by \citet{shen_2007}, who analysed three categories of grouped decisions: location-routing, inventory-routing, and location-inventory. \citet{klibi_et_al_2010} reviewed robust supply chains under uncertainty, while \cite{farahani_et_al_2014b} analysed the SCND literature. 

We propose the classification of healthcare mathematical models following the aims of the healthcare supply chains, namely maximising availability, accessibility, and adaptability \citep{daskin_dean_2004}. Availability addresses the various uncertain parameters such as manufacturing process duration or demand. Any such uncertainty can lead to disruptions and the need to make further changes to the scheduling of a patient or product. Accessibility aims to maximise the covered demand under given constraints related to time, distance, or cost. Adaptability measures the level to which adjustments are possible, given the need of fundamental changes due to later events \citep{eckstein_et_al_2014}. Such instances include closure of a hospital or manufacturing facility and create disruptions at the strategic level of the network. 

The two most widely used facility location models in non-emergency healthcare are Location Set Covering Problem (LSCP) and Maximal Covering Location Problem (MCLP). The first was initially proposed by \citet{toregas_et_al_1971} and aims to minimise the total number of facilities needed so that the entire demand is covered by at least one facility. \citet{vasko_wilson_1984} extended the problem by adding non-uniform facility costs and a constraint that limits the distance between a facility and a demand point. The MCLP was proposed by \cite{church_revelle_1974} and aims to maximise the covered demand by imposing distance or time constraints and a limited number of facilities to be located.

The availability aspect was also addressed by \cite{hogan_revelle_1986}, who proposed two formulations that consider a backup coverage in case of disruptions. In their paper the objectives ensure that each demand point is covered by at least two facilities. The formulation has a hierarchical construction through which the first, or required, coverage distance is smaller than the one for second, or backup, coverage. In this case, both distances are constrained to an upper, predefined threshold. The backup coverage models can be of special aid for ATMPs where, constrained by the number of available skilled labour, it might be beneficial to use sparser facilities instead of an overly centralised approach. However, solving these models can quickly become computationally expensive for large decision spaces.

Nevertheless, \citet{murray_et_al_2010} highlights the fact that both LSPC and MCLP formulations assume that the facilities are available without interruptions. However, idle periods for some facilities are frequent in many supply chains, including PM due to the very specialised type of machines used for the manufacturing of ATMPs. suggested the Implicit and Explicit formulations of the classic models. In this case, the demand node represents an area rather than a single point; the set of candidate facility locations can cover an area; and the decision variables measure whether an area is sufficiently covered by a facility. The implicit models assume that an area can be covered by using multiple facilities at different geographical points. The explicit models work with different configurations of fixed supply facilities and calculate the exhaustive number of coverage levels for a demand area. The coverage levels represent a predefined value denoting the maximum number of facilities allowed to partially cover an area.

Similar concerns have also been addressed in the past by formulations such as the Hierarchical Objective Set Covering Model (HOSC) by \cite{daskin_stern_1981}; and probabilistic versions by \cite{revelle_hogan_1989b} or Queuing Probabilistic Location Set Covering Problem (QPLSCP) by \cite{marianov_revelle_1996}.

The accessibility component of the healthcare supply chains is most commonly solved using location-allocation formulations. While many extensions have been proposed, the basic terminologies are inherent to the classical p-median and p-center formulations \citep{rahman_smith_2000, dzator_dzator_2017}. Constructed as minmax problems, the common objective is minimising the total transportation time or distance between the supply and demand points. The constraints ensure perfect demand coverage, allocation only to open facilities, and an exact number of facilities to be placed. The difference between the two formulations comes from the restriction of maximum allowed time or distance between two nodes.

For development-stage companies, the strategic level decisions, such as facilities location, are most often taken during times of high levels of uncertainty like commercialisation approvals, and adaptability is a key component in SCM. Scenario planning is one of the approaches commonly used in healthcare to mitigate this variability. \citet{daskin_dean_2004} were the first to formulate an extension to the p-median problem by considering different scenarios. Three objectives are commonly followed: minimise the total/maximum distance or time for the entire demand over all scenarios and minimise the maximum regret over scenarios starting from an optimal predefined value.

\subsection{Proposed Personalised Medicine Model} \label{sec:personalised_model}

The modelling work in this paper incorporates the view of a central decision maker who aims to find optimal locations for the required facilities and assign the patient's order. As part of the PM supply chain, manufacturing facilities will process the ATMP, while cryopreservation facilities are responsible with the temperature sensitivity and shelf-life constraints. The network configuration should lead to a low cost and low waiting time, and ensure that as many patients as possible can be offered the ATMP. Unlike the first two objectives which are common in the healthcare supply chain literature, the third objective is more commonly treated as a constraint. However, the current personalised ATMPs are not currently available at all hospitals. In our case, we do not ensure perfect coverage since we start the supply chain from the hospital and assume that patients could travel to the hospitals that can offer ATMPs. 

The time and cost in our model are dependant on the mode of production used. A mode refers to the sum of manufacturing steps necessary to produce an ATMP from cell collection and until administration. New manufacturing modes are constantly developed by biopharmaceutical companies since the optimization of the manufacturing process is an active research area. To exemplify what different modes look like, we use the classification created by \citet{lopes_et_al_2020} into fully manual, semi-automated, and fully automated manufacturing. The three production modes differ in their reliance on automated systems and skilled labour resources. For instance, a fully automated process can increase the production capacity per facility, reduce the time and reliance on skilled labour. Nonetheless, the high investment in automated systems required to operate such a facility might not be profitable in areas with very low demand which is characteristic to PM products.

The production modes are also linked to the possibility of damaging the cells and can lead to product failure \citep{lopes_et_al_2020}. If an order does not pass the quality checks, it cannot be returned to the patient and the entire supply chain is restarted. In our model, we thus assume a corresponding failure rate to each production mode and increment the cost and the waiting time accordingly. This failure rate can, however, be extended in future models to account for patient's health condition which can affect the quality of the starting product. 
\begin{figure}[hbt!]
\noindent\textbf{Indices and Parameters}
\newline\nopagebreak%
{\renewcommand{\arraystretch}{1}%
  \begin{tabular}{r@{\hskip 1em}p{\textwidth - 5em}@{}}
   $i \in I$ & Demand nodes (individual orders).\\
   $j \in J$ & Candidate facility locations.\\
   $k \in K$ & Manufacturing  modes  (e.g., manual, semi-automated, automated). \\
   $d_{ij}$ & Travel time between (facility or demand node) locations  $i,j \in I \cup J$.\\
   $s^\text{M}_j, s^\text{C}_j$& Setup cost (e.g., construction) of a facility at location $j \in J$.\\
   $c^\text{f}_{ijk}, c^\text{z}_{ijk}$& Operation cost per fresh (f) and frozen (z) order covered that depends on the order $i$, location $j$ and the mode $k$. \\
   $p^\text{f}_k, p^\text{z}_k$ & Production time per fresh (f) and frozen (z) order for mode $k \in K$.  \\
   $r_{ik}$ & Failure rate, within $[0,1]$,  of manufacturing for order $i \in I$ when using  mode $k$. \\
   $\gamma_i$ & Shelf-life of order $i \in I$ when processed as fresh cells. \\
   $T$ & Constant  larger than any travel time. \\
\end{tabular}}
\end{figure}

\begin{figure}[hbt!]
\noindent\textbf{Decision Variables}
\newline\nopagebreak
{\renewcommand{\arraystretch}{1}
  \begin{tabular}{@{}r@{\hskip 1em}p{\textwidth - 5em}@{}}
     $y^\text{M}_j$, $y^\text{C}_j$ & 1 if a manufacturing (M) or cryo-preservation (C) facility is opened at candidate location $j \in J$; 0 otherwise. \\
     $x^\text{M}_{ij}$, $x^\text{C}_{ij}$ & 1 if demand location $i \in I$ is covered by the manufacturing (M) or cryo-preservation (C) facility at candidate location $j \in J$; 0 otherwise. \\
     $z_i$ & 1 if demand order $i \in I$ is cryo-preserved; 0 otherwise. \\
     $m_{jk}$ & 1 if facility at location $j \in J$ is assigned a manufacturing mode  $k \in K$; 0 otherwise. \\
\end{tabular}} 
\end{figure}

\vspace{-1em}

\begin{table}[hbt!]
  \newcommand{\tabeqlabel}[1]{\refstepcounter{tableeqn}(\thetableeqn)\label{#1}}
  \newcommand\mcol\multicolumn
    \stepcounter{table}
\centering
\begin{tabular}{@{}rlrr@{\hskip 2em}lrr@{}}
\multicolumn{6}{@{}l}{\textbf{Personalised Medicine Supply Chain Model}}\\
Objectives  
            & \mcol{2}{l}{min \ \textit{W} } & \tabeqlabel{pm.obj1} \\
            & \mcol{2}{l}{min \ \textit{C}}& \tabeqlabel{pm.obj2}\\
            & \mcol{2}{l}{max \ \textit{V}}&\tabeqlabel{pm.obj3}\\
      Constraints 
            & {$\displaystyle\sum_{j \in J} x^\text{M}_{ij} \cdot d_{ij} \leq \gamma_i + z_i \cdot T$} &$ \forall  i \in I$ &\tabeqlabel{pm:fresh_to}
            
            & {$\displaystyle\sum_{j \in J} x^\text{C}_{ij} d_{ij} \leq 24\,\text{hours}$}& $ \forall i \in I$ & \tabeqlabel{pm:frozen}  \\
            
            & {$\displaystyle\sum_{j \in J} x^\text{M}_{ij} \cdot d_{ji} \leq \gamma_i + z_i \cdot T$} &$ \forall i \in I$&\tabeqlabel{pm:fresh_from}
            
            & $\displaystyle\sum_{j \in J} x^\text{M}_{ij} \leq 1$ &$ \forall i \in I$ &\tabeqlabel{pm:one_manufacturing}\\
            
            & $\displaystyle\sum_{j \in J} x^\text{C}_{ij} \leq z_i$&$ \forall i \in I$ &\tabeqlabel{pm:one_cryo}
            
            & \mcol{2}{l}{$x^\text{M}_{ij} \leq y^\text{M}_{j} \hspace*{\fill} \forall i \in I, j \in J$} &\tabeqlabel{pm:m_assign_ifopen}\\  
            
            & \mcol{2}{l}{$x^\text{C}_{ij} \leq y^\text{C}_{j}  \hspace*{\fill}  \forall i \in I, j \in J$} &\tabeqlabel{pm:c_assign_ifopen}
            
            & $y^\text{M}_{j} + y^\text{C}_{j} \leq 1 $&$\forall j \in J$ &\tabeqlabel{pm:one_type}\\ 
            
            & $\displaystyle\sum_{k \in K} m_{jk} = y^\text{M}_j $&$ \forall j \in J$ &\tabeqlabel{pm:mode_manufacturing}\\
            
            & \mcol{5}{r}{$x^\text{M}_{ij},\  x^\text{C}_{ij},\ y^\text{M}_{j},\ y^\text{C}_{j}, z_i,\ m_{jk} \in \{0, 1\}$} &\tabeqlabel{pm:ranges}\\
\end{tabular}
\end{table}

\begin{figure}[hbt!]
    \centering
    \includegraphics[width=\textwidth]{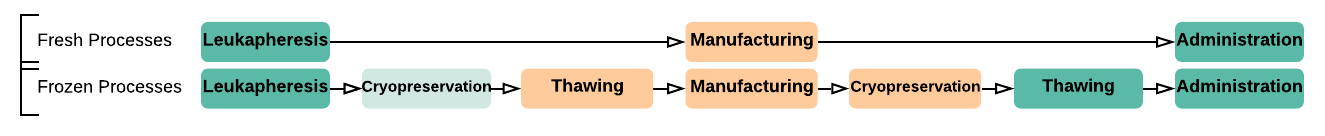}
    \caption{Differences in the supply chain between fresh and frozen processes.}
    \label{fig:production_processes}
\end{figure}

\paragraph{Objective 1: Waiting Time.} The waiting time is the time that a patient (represented by order $i \in I$) must wait from their cells being collected (leukapheresis) until the ATMP is administered. Depending on whether the order is processed fresh or frozen, the steps are slightly different, as shown in Fig.~\ref{fig:production_processes}.

Let us assume, for simplicity, that leukapheresis time is already included in all travel times starting from each demand location $i \in I$. Similarly, administration time is included in the travel times to each demand location. Then, for an order that is fresh (f), the waiting time corresponds to the travel time from/to the demand location $i \in I$ and its assigned manufacturing facility in location $j \in J$ plus its production time, which depends on the manufacturing mode $k \in K$,
whereas if the order is cryopreserved, we first visit a cryopreservation facility at location $j'$ and then a manufacturing facility at location $j$.

Therefore, by taking into account whether order $i$ is cryopreserved ($z_i$) or not, the waiting time for order $i$ can be expressed as:
\begin{equation}
  \label{eq:waiting_time}
  \sum_{j \in J} x^\text{M}_{ij} \Bigl [%
  \underbrace{(1-z_i)  d_{ij} 
    +  z_i\Bigl( \sum_{j'\in J} x^\text{C}_{ij'} (d_{ij'} + d_{j'j})\Bigr)}_{\textit{TimeToManufacture}(i,j)} %
  + \underbrace{\sum_{k\in K} \Bigr((1 - z_i)p^\text{f}_k + z_i p^\text{z}_k \Bigl) m_{jk}}_{\textit{ProductionTime}(i,j)} %
  + \underbrace{\phantom{\sum_j^J} d_{ji}\phantom{\sum_j^J}}_{\textit{TimeToPatient}(i,j)}\Bigr]
\end{equation}

The failure rate gives the frequency that the production of an order may fail, thus the process needs to repeated again from the leukapheresis up to completing production an average of $(1 + x^\text{M}_{ij} \sum_{k\in K} r_{ik} m_{jk})$ times, thus the total waiting time is calculated as:

  \begin{multline}
    \textit{W} = \sum_{i \in I} \sum_{j \in J} x^\text{M}_{ij} \Bigl[  \bigl(1 + \sum_{k\in K} r_{ik}  m_{jk}\bigr) \bigl(\textit{TimeToManufacture}(i,j)+ \textit{ProductionTime}(i,j)\bigr) \\
    + \textit{TimeToPatient}(i,j) \Bigr] 
  \end{multline}

 \paragraph{Objective 2: Cost.}

Each facility type has a setup cost $s^\text{M}_j, s^\text{C}_j$ that depends on the location and an operation cost $c^\text{f}_{ijk}, c^\text{z}_{ijk}$ that depends on the order $i$, the location $j$, the manufacturing mode $k$, and whether an order is fresh (f) or frozen (z). It already takes into account resources and transportation for each order. Thus, total cost (without considering failures) of the supply chain is given by:
\begin{equation}
  \label{eq:cost}
  \sum_{j \in J}  \bigl( s^\text{M}_j y^\text{M}_j + s^\text{C}_j y^\text{C}_j \big) + \sum_{i \in I} \sum_{j \in J} \sum_{k \in K} x^\text{M}_{ij} m_{jk} \bigl((1-z_i)c^\text{f}_{ijk}  + z_i c^\text{z}_{ijk}  \bigr)
\end{equation}

If manufacturing fails (see ``Reject'' in Fig.~\ref{fig:supply_chain_general}), we need to process the complete order again, which can be modelled as a factor that increments the operation costs:
\begin{equation}
  \label{eq:TotalCost}
  \textit{C} = \sum_{j \in J}  ( s^\text{M}_j y^\text{M}_j + s^\text{C}_j y^\text{C}_j )  + \sum_{i \in I} \sum_{j \in J} x^\text{M}_{ij} \sum_{k \in K}\bigr(1 +   r_{ik} \bigl) m_{jk} \bigl((1-z_i)c^\text{f}_{ijk}  + z_i c^\text{z}_{ijk}  \bigr)
\end{equation}

 \paragraph{Objective 3: Coverage.}
Since each order must be fulfilled by one manufacturing facility, the total
coverage, that is, the number of demand orders that can be covered or, in other
words, the number of patients that can receive the therapy is given by:
\begin{equation}
  \textit{V} = \sum_{i\in I}\sum_{j \in J} x^\text{M}_{ij}
  \label{eq:coverage}
\end{equation}

\paragraph{Constraints.}

The most important constraints concern the shelf-life $\gamma_i$, typically no more than a few days, of the ATMP corresponding to order $i \in I$, which represents the maximum travel time either to (\ref{pm:fresh_to}) or from (\ref{pm:fresh_from}) the manufacturing facility.
The shelf-life to travel to a cryopreservation facility is more strict (preferably no more than 24 hours, which is always less than $\gamma_i$) to compensate for the damage caused by the freezing process (\ref{pm:frozen}). After cryopreservation, shelf-life is unlimited for all practical purposes, hence, we disable constraints (\ref{pm:fresh_to}) and (\ref{pm:fresh_from}) using a very large shelf-life $T$ if the order is cryopreserved ($z_i = 1$).

Other constraints ensure that each order is assigned to no more than one manufacturing facility (\ref{pm:one_manufacturing}), each order is assigned to no more than one cryopreservation facility only if the order is cryopreserved (\ref{pm:one_cryo}), and order is assigned to a facility at a given location if a facility of that type is open at that location (\ref{pm:m_assign_ifopen}) and (\ref{pm:c_assign_ifopen}), only one facility type may be opened at the same location (\ref{pm:one_type}), and one, and only one, mode is assigned at each location, and only if there is an open manufacturing facility at that location (\ref{pm:mode_manufacturing}).

\section{Conclusion and Future Research} \label{sec:conclusion}

The treatments developed under PM have led to the need for an inclusive supply chain in the healthcare sector. The timely and cost-efficient delivery of biopharmaceutical products of human origin are challenged by aspects that have been researched only independently in the past. Our growing ability to develop ATMPs increases the need for a new cold supply chain which accounts for the delivery, as well as the manufacturing aspects. Furthermore, a new element is also brought by the concepts of "living products" and the "process is product" where the final ATMP is defined by the entire supply chain network and its configuration.

This paper presented a first systematic review of literature from an operational research perspective for the ATMPs supply chain in light of the current knowledge in the medical and pharmaceutical sectors. With the aim of understanding the exact dissimilarities between the targeted sectors and ATMPs, the extensive literature mapping concentrated on the recent publications in the traditional non-emergency healthcare networks from a modelling perspective. Our analysis suggests that, while the PM distribution is sharing common characteristics with other mature supply chains, none of these mirrors the PM network. The existing models in the literature were designed for mass delivery as a result of high demand and, as a consequence, are not appropriate to cover some of the key challenges of PM. 

The bibliographic and modelling analysis presented in this review indicates that, while the means for PM optimisation exist, more comprehensive research is necessary. We proposed a new mathematical model that integrates the location of multiple facility types and the allocation of demand from hospitals. Three objectives are considered, namely cost minimisation (i.e. construction, manufacturing, and delivery costs), time minimisation (i.e. delivery and manufacturing duration), and demand coverage maximisation. Our model integrates the manufacturing process alongside logistics and considers multiple production modes that influence the total cost and time of the supply chain, and each order's success rate. The model is constrained by the most important PM characteristic, i.e. short shelf-life, which dictates whether an order needs to be cryopreserved or not.

Our understanding of the commercialisation of PM products from an operations research perspective is still incipient and further research is deemed necessary. For instance, future mathematical models for PM can look at understanding how patient prioritisation and risk analysis can lead to an equitable and fair supply chain. In this way, we ensure that the entire process is designed for each individual and that the most at risk patients can have access to timely personalised medical therapies. 

With respect to solution methods, most of the studies analysed in this review used exact methods to solve the problems to optimality. Nevertheless, a real-world scenario for ATMPs is driven by highly uncertain parameters and the need to make the product available to a global market makes this approach infeasible. Additionally, the biopharmaceutical companies might not have access to the commercial solvers used in the papers and a transposal of the corresponding algorithm can become challenging. This is something that should be borne in mind in future studies.

\bibliographystyle{unsrtnat}

\end{document}